\documentclass[useAMS,usenatbib]{mn2e}

\usepackage{graphicx}
\usepackage{natbib}

\newcommand{\ltsima} {$\; \buildrel < \over \sim \;$}
\newcommand{\gtsima} {$\; \buildrel > \over \sim \;$}
\newcommand{\lta} {\lower.5ex\hbox{\ltsima}}
\newcommand{\gta} {\lower.5ex\hbox{\gtsima}}

\title[The effect of the linear term on the $f_{nl}$ wavelet
  estimator]{The effect of the linear term on the wavelet estimator of
  primordial non-Gaussianity}

\author[A. Curto et al.]{A. Curto, \thanks{e-mail:
curto@ifca.unican.es} E. Mart\'inez-Gonz\'alez, R. B. Barreiro \\
     Instituto de F\'isica de Cantabria, CSIC-Universidad de Cantabria, Avda. de los Castros s/n, 39005 Santander, Spain.\\
}
\date{Accepted  Received ; in original form }
\begin{document}
\maketitle
\begin{abstract}
In this work we present constraints on different shapes of primordial
non-Gaussianity using the Wilkinson Microwave Anisotropy Probe (WMAP)
7-year data and the spherical Mexican hat wavelet $ f_{nl}$ estimator
including the linear term correction. In particular we focus on the
local, equilateral and orthogonal shapes. We first analyse the main
statistical properties of the wavelet estimator and show the
conditions to reach optimality.  We include the linear term correction
in our estimators and compare the estimates with the values already
published using only the cubic term. The estimators are tested with
realistic WMAP simulations with anisotropic noise and the WMAP $KQ75$
sky cut. The inclusion of the linear term correction shows a
negligible improvement ($\le$ 1 per cent) in the error bar for any of
the shapes considered. The results of this analysis show that, in the
particular case of the wavelet estimator, the optimality for WMAP
anisotropy levels is basically achieved with the mean subtraction and
in practical terms there is no need of including a linear term once
the mean has been subtracted. Our best estimates are now: $
\hat{f}^{(loc)}_{nl}= 39.0 \pm 21.4$, $\hat{f}^{(eq)}_{nl}= -62.8 \pm
154.0$ and $\hat{f}^{(ort)}_{nl}= -159.8 \pm 115.1$. We have also
computed the expected linear term correction for simulated Planck maps
with anisotropic noise at 143 GHz following the Planck Sky Model and
including a mask. The improvement achieved in this case for the local
$f_{nl}$ error bar is also negligible (0.4 per cent).
\end{abstract}
\section{Introduction }
\label{sec:introduction}
In the recent years the spherical Mexican hat wavelet (SMHW)
\citep{martinez2002} has been used to construct a new type of
estimator for the primordial non-Gaussianity in the CMB characterised
by the non-linear coupling parameter $f_{nl}$
\citep{curto2009a,curto2009b,curto2010,curto2011a,curto2011b}. One of
the particularities of the wavelet estimator as it has been
traditionally presented in the literature compared with direct
bispectrum-based estimators
\citep{komatsu2001,komatsu2002,komatsu2003,babich2004,babich2005,creminelli2006,creminelli2007,wandelt2008,smith2009,elsner2009,liguori2010,senatore2010,smidt2010,fergusson2010a,fergusson2010b,komatsu2011,fergusson2011}
is the absence of a linear term. In the bispectrum-based estimators,
the linear term plays a key role to achieve optimality in the cases
where the rotational invariance of the CMB is broken because of
different instrumental complexities such as anisotropic noise or
partial sky coverage \citep[see for
  example][]{creminelli2006,creminelli2007,yadav2010,fergusson2011}.

The computational difficulties related to the inversion of the
covariance matrix present in the bispectrum estimator, especially in
future data sets with higher $\ell_{max}$ as for example
Planck\footnote{http://www.esa.int/planck}, together with the unknown
effect that different systematics from the instrument and background
residuals might have on the estimates, motivated the search for new
estimators based on different tools such as the SMHW described in
this paper, the binned bispectrum \citep{bucher2010}, the general
modal expansion and polyspectra estimation
\citep{fergusson2010b,fergusson2011}, the needlets
\citep{marinucci2008,pietrobon2009,rudjord2009,donzelli2012}, the
HEALPix wavelet \citep{casaponsa2011a}, neural networks
\citep{casaponsa2011b} or a Bayesian approach
\citep{elsner2010a,elsner2010b} among others.

In a previous paper \citep{curto2011a}, we described the main features
of the wavelet estimator based on the cubic statistics constructed
from the SMHW coefficient maps. Those cubic terms were written as a
function of the non-linear coupling parameter $f_{nl}$ and the
bispectrum of the primordial non-Gaussianity.  In that paper we also
showed that the power of the method to detect $f_{nl}$, that is the
variance of this parameter $\sigma^2(f_{nl})$, matches that of the
direct bispectrum-based estimators for ideal conditions (full sky and
isotropic noise) and realistic conditions (partial sky coverage and
anisotropic noise). The wavelet estimator variance was obtained in two
different ways: through the Fisher matrix and by means of Monte Carlo
(MC) simulations, providing very similar results. A remarkable result
of these works is the fact that the wavelet estimator is, in practice,
able to reach optimality on the $f_{nl}$ estimation without including
any linear term correction. However, from several works \citep[see for
  example][]{creminelli2006,fergusson2011} it has been shown that in
order to reach minimum variance, all the cubic estimators need a
linear term correction. A recent work has solved this apparent
controversy \citep{donzelli2012} by showing that in
WMAP\footnote{http://map.gsfc.nasa.gov/} anisotropy conditions, the
linear term correction is nearly equivalent to the mean subtraction
performed for each wavelet coefficient map in the wavelet estimator.

In this paper we re-examine the main statistical properties of the
wavelet estimator and show the conditions to reach optimality. We
compute the linear term correction for the local, equilateral and
orthogonal $f_{nl}$ shapes. In particular we see that the linear term
correction for the local case provides a 1 per cent reduction in the
error bars \citep[in agreement with ][ for the SMHW]{donzelli2012}
while the correction for the other shapes is even smaller. Section
\ref{sec:estimator} introduces the SMHW estimator, its variance and
its linear correction. In Section \ref{sec:wmap_data} the estimator
with its linear correction is applied to WMAP 7-year data for the
local, equilateral and orthogonal shape. In Section
\ref{sec:planck_simulations} we explore the linear correction on
Planck simulations at 143 GHz for the local shape, and in Section
\ref{conclusions} the conclusions are presented.

\section{The wavelet approach}
\label{sec:estimator}
In this Section we present an approach for the $f_{nl}$ estimator
based on the statistical properties of the cubic terms of the SMHW
coefficients averaged over the sky.  In this case we exploit the
property of the SMHW wavelet that performs a strong decorrelation of
the data at distances larger than the wavelet resolution. The expected
value of the cubic terms in the sky are obtained from the sum of a
large number of almost independent elements and therefore its
distribution will be close to Gaussian by the central limit
theorem. We will first review the SMHW and its decorrelation
properties and then we will construct the wavelet estimator based on
those properties including the linear term correction.
\subsection{The SMHW coefficients and their correlation}
\begin{figure}
\centering
\includegraphics[width=7.1cm, height=5.cm, angle= 0]{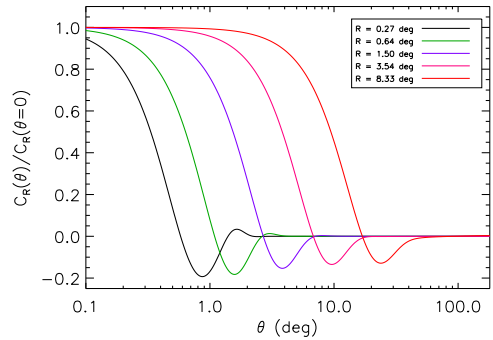}
\caption{The normalised correlation functions for different wavelet
  coefficient maps are plotted. From {\it left to right}, the curves
  correspond to maps convolved with a SMHW of scale R = 0.27, 0.64,
  1.50, 3.54 and 8.33 degrees, respectively. \label{fig_corr_smhw}}
\end{figure}
Detailed information about the spherical Mexican Hat wavelet (SMHW)
and a (non-complete) list of applications to the CMB maps and cosmology
can be found in
\citet{antoine1998,martinez2002,cayon2003,vielva2006,vielva2007b,mcewen2007,martinez2008,zang2011,yu2012}.
      
Given a function $f(\bf n)$ defined at a position $\bf n$ on the sphere and
a continuous wavelet family on that space $\Psi({\bf n}; {\bf b}, R)$, we
define the continuous wavelet transform as
\begin{equation}
w(R; {\bf b}) = \int d{\bf n}f({\bf n})\Psi({\bf n}; {\bf b}, R)
\label{wavmap}
\end{equation}
where ${\bf b}$ is the position on the sky at which the wavelet
coefficient is evaluated, $R$ is the scale of the wavelet and $\Psi_S(\theta; R) \equiv \Psi({\bf n(\theta, \phi)}; {\bf 0}, R)$ is
given by
\begin{equation}
\Psi_S(\theta; R) = \frac{1}{\sqrt{2\pi}N(R)}\left [ 1+ \left( \frac{y}{2}\right)^2\right ]^2\left [ 2 - \left( \frac{y}{R}\right)^2\right ]e^{-y^2/(2R^2)}
\end{equation}
where
\begin{equation}
N(R)=R\left(1+\frac{R^2}{2}+\frac{R^4}{4}\right)^{1/2}
\end{equation}
and
\begin{equation}
y = 2\tan\left( \frac{\theta}{2}\right).
\end{equation} 
Considering a set of different angular scales $\{R_i\}$ we define a
third order statistic depending on three scales $\{i,j,k\}$
\citep{curto2009b}
\begin{equation}
q_{i j k}=\frac{1}{4\pi}\frac{1}{\sigma_i\sigma_j\sigma_k}\int d{\bf n} w(R_i,{\bf n})w(R_j,{\bf n})w(R_k,{\bf n})
\label{themoments_qijk}
\end{equation}
where $\sigma_i$ is the dispersion of the wavelet coefficient map
$w(R_i,{\bf n})$. In the particular case of $R_0=0$, $w(R_0,{\bf n})
\equiv f({\bf n})$. For a particular pixelization on the sphere,
Eq. \ref{themoments_qijk} can be written as:
\begin{equation}
\label{statistic}
\bf q_{i j k}=\frac{1}{N_{ijk}}\sum_{p=0}^{N_{pix}-1}\frac{w_{i}(p)w_{j}(p)w_{k}(p)}{\sigma_i\sigma_j\sigma_k}
\end{equation}
where $ N_{pix}$ is the total number of pixels of the map, $ N_{ijk}$
is the number of pixels available after combining the extended masks
corresponding to the three scales $ R_i$, $ R_j$ and $ R_k$ and $
w_{i}(p) \equiv w(R_i,p) - \langle w(R_i) \rangle$ is the wavelet
coefficient in the pixel $ p$ evaluated at the scale $ R_i$ after
subtracting the mean value over the wavelet coefficient map outside
its extended mask.

Using the properties of the wavelet, we may write the wavelet
  transform of the temperature map in the next form
\citep{curto2011a}
\begin{equation}
w(R_i,{\bf n})=\sum_{\ell m}a_{\ell m}\omega_\ell (R_i)Y_{\ell m}(\bf n).
\label{thewaveletconv}
\end{equation}
Using the isotropic properties of the CMB and the properties of the
wavelet, we can obtain the angular coefficient correlation
$C_{ij}(\theta)$ between any pair of pixels ${\bf n}$ and ${\bf n'}$
separated by an angular distance ${\bf n}{\bf n'}= \cos(\theta)$ and
for two angular scales $R_i$ and $R_j$
\begin{eqnarray}
\nonumber
& C_{ij}(\theta)  \equiv  \langle  w(R_i,  {\bf n})w(R_j,  {\bf  n})  \rangle = & \\
& = \sum_{\ell}\frac{2\ell+1}{4\pi}C_{\ell}\omega_\ell(R_i)\omega_\ell(R_j)P_{\ell}
\left(\cos(\theta)\right) &
\label{thewaveletdisp}
\end{eqnarray}
where $\omega_\ell (R)$ is the window function of the wavelet at a
scale $R$ and it is given by the harmonic transform of the mother
wavelet of the SMHW \citep{martinez2002,sanz2006}. The dispersion of
the wavelet coefficients at scale $R_i$ is simply given by
$\sigma_i = C_{ii}(\theta=0)^{1/2}$.

In Fig. \ref{fig_corr_smhw} we show the correlation of the wavelet
coefficients as a function of the angular distance $\theta$ for
several values of the resolution scale $R$. As can be seen, the SMHW
produce an effective decorrelation of the signal at angular distances
above the resolution scale $R$.
\subsection{The wavelet estimator}
\begin{figure*}
\begin{center}
\includegraphics[width=4.0cm,height=3.0cm, angle = 0] {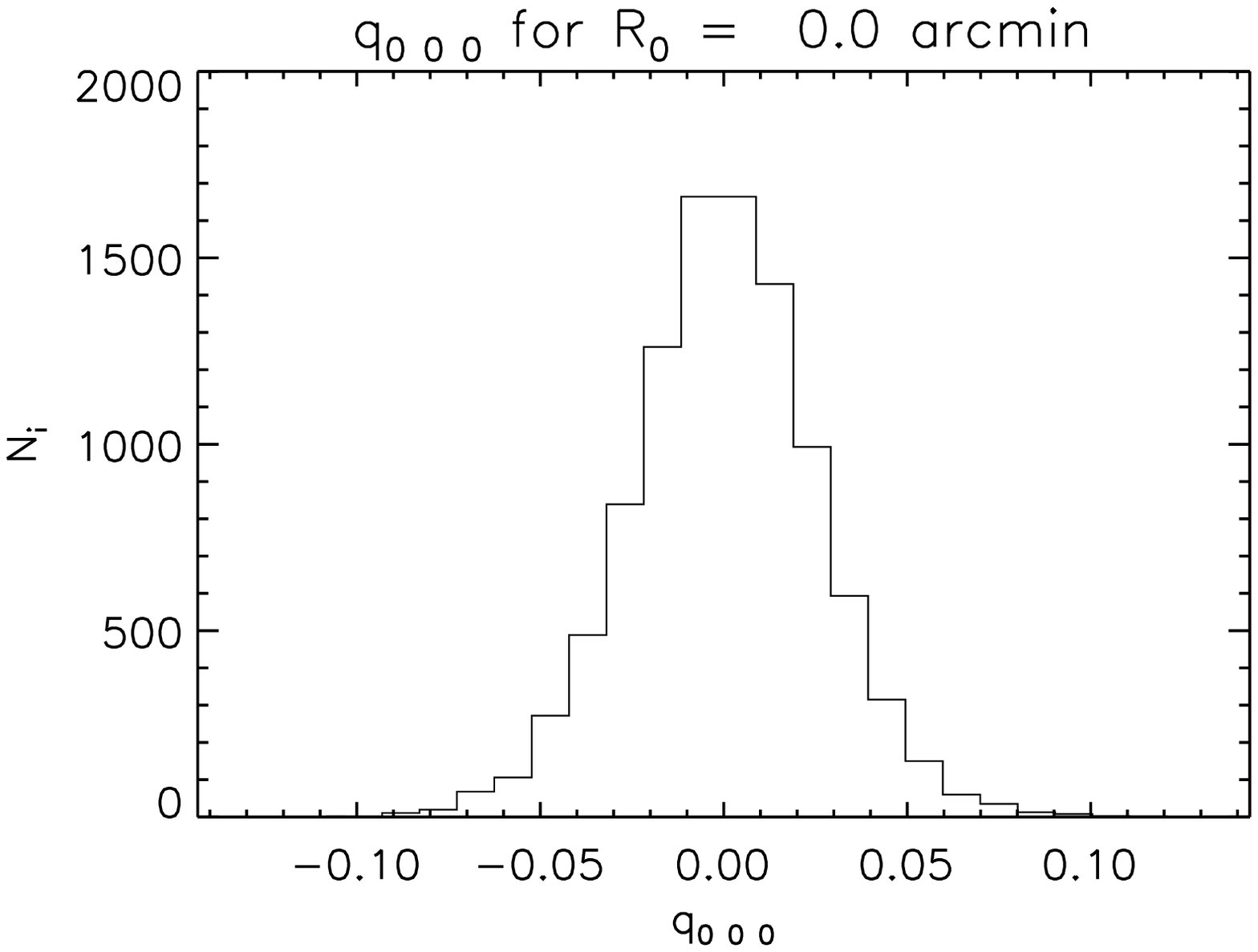}
\includegraphics[width=4.0cm,height=3.0cm, angle = 0] {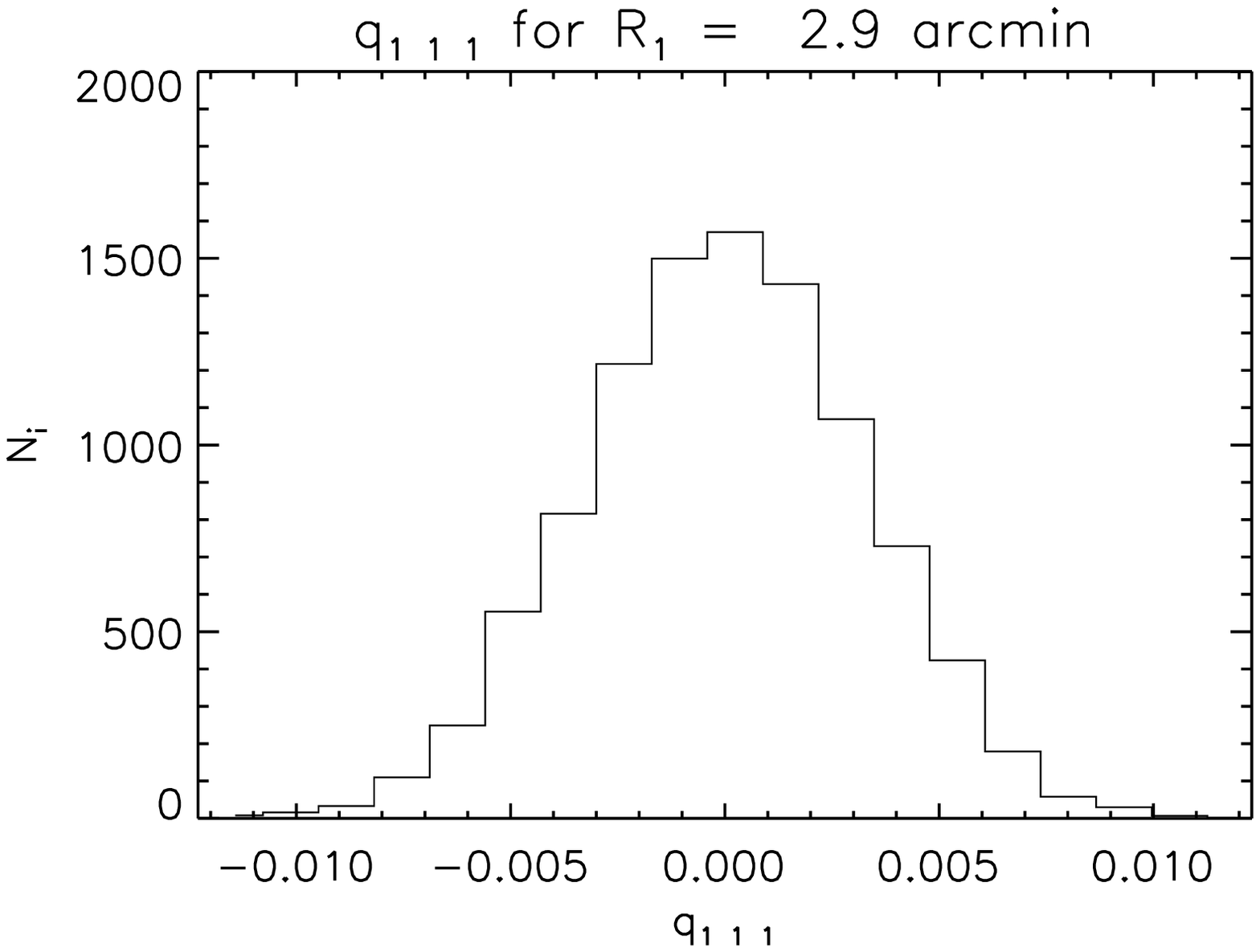}
\includegraphics[width=4.0cm,height=3.0cm, angle = 0] {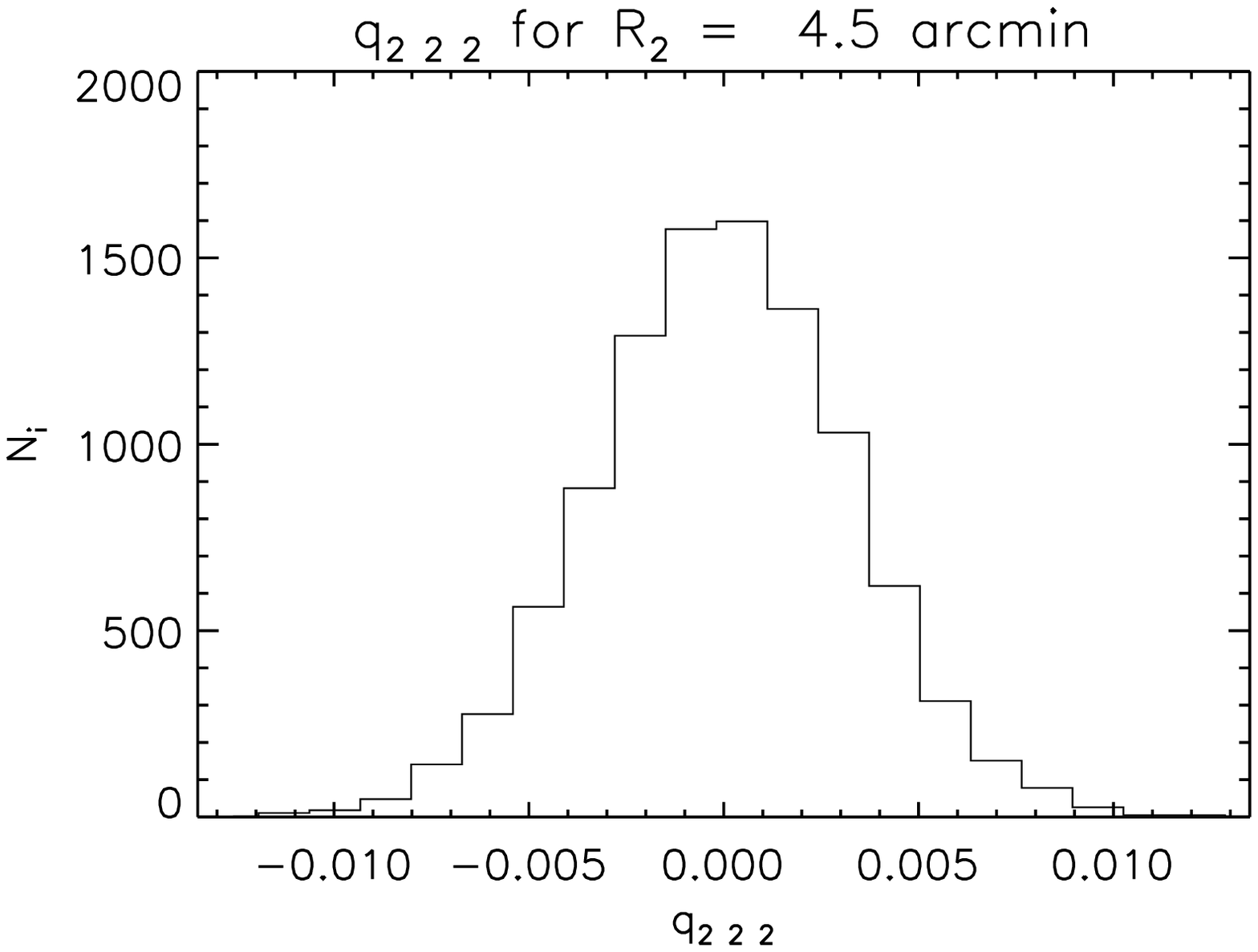}
\includegraphics[width=4.0cm,height=3.0cm, angle = 0] {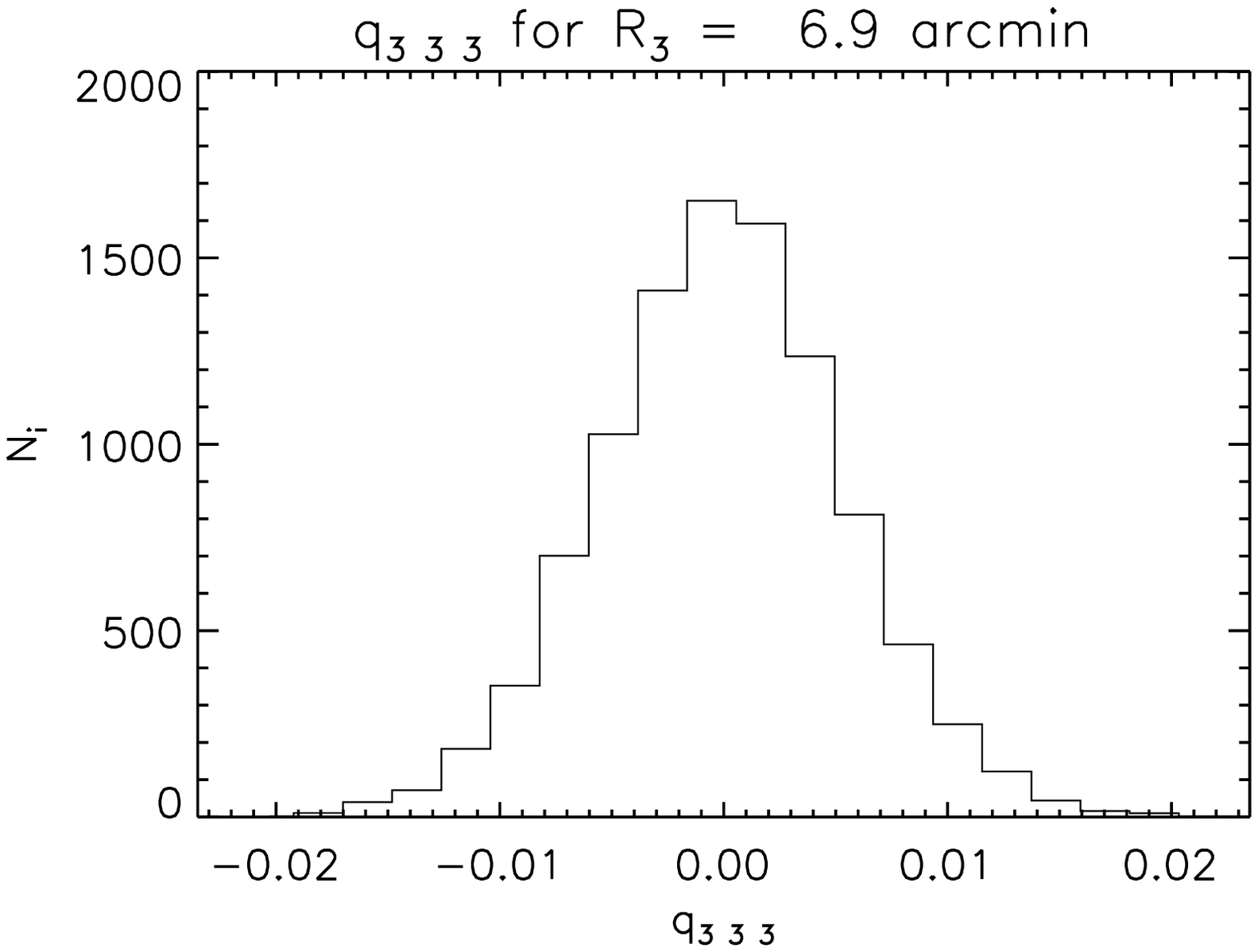}
\includegraphics[width=4.0cm,height=3.0cm, angle = 0] {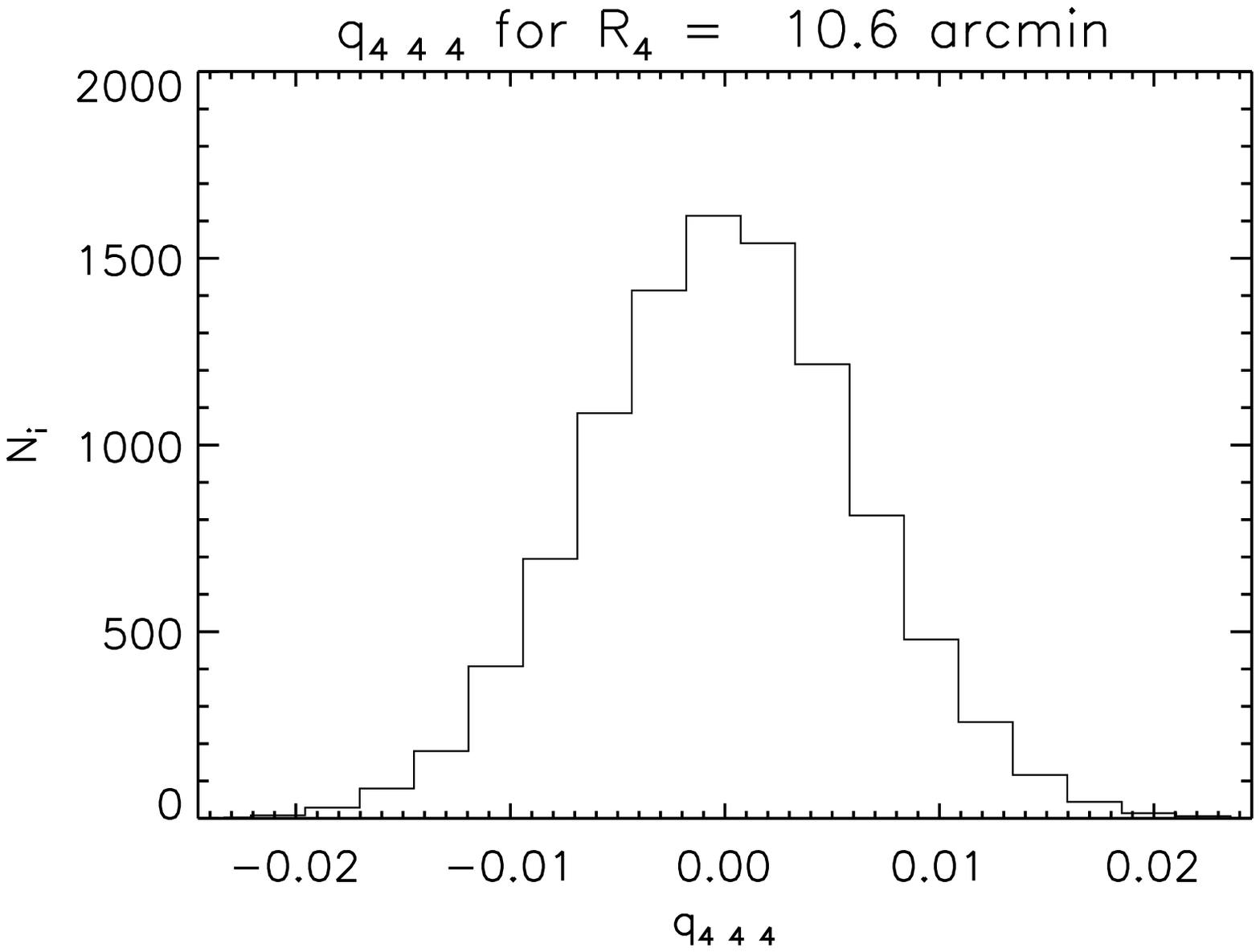}
\includegraphics[width=4.0cm,height=3.0cm, angle = 0] {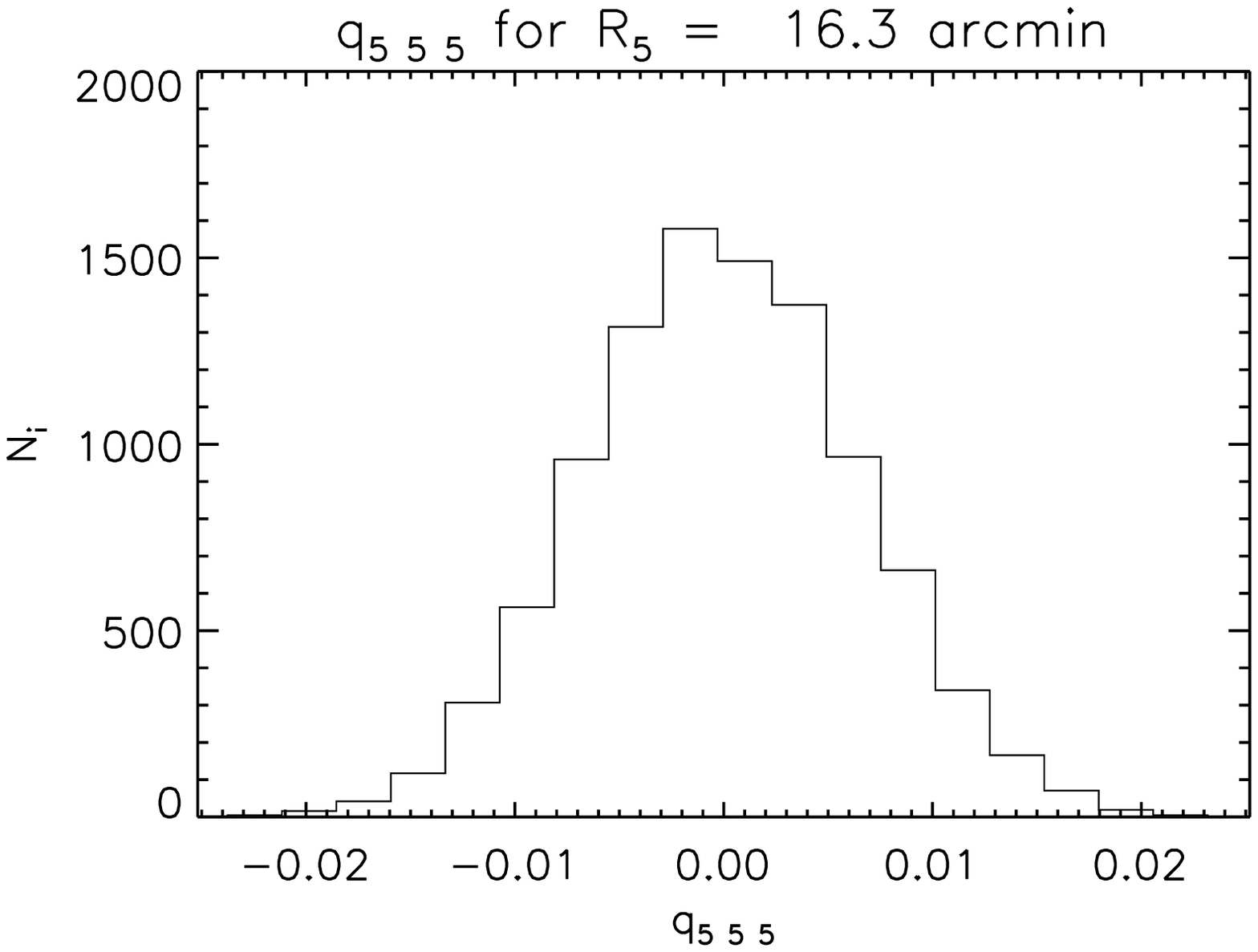}
\includegraphics[width=4.0cm,height=3.0cm, angle = 0] {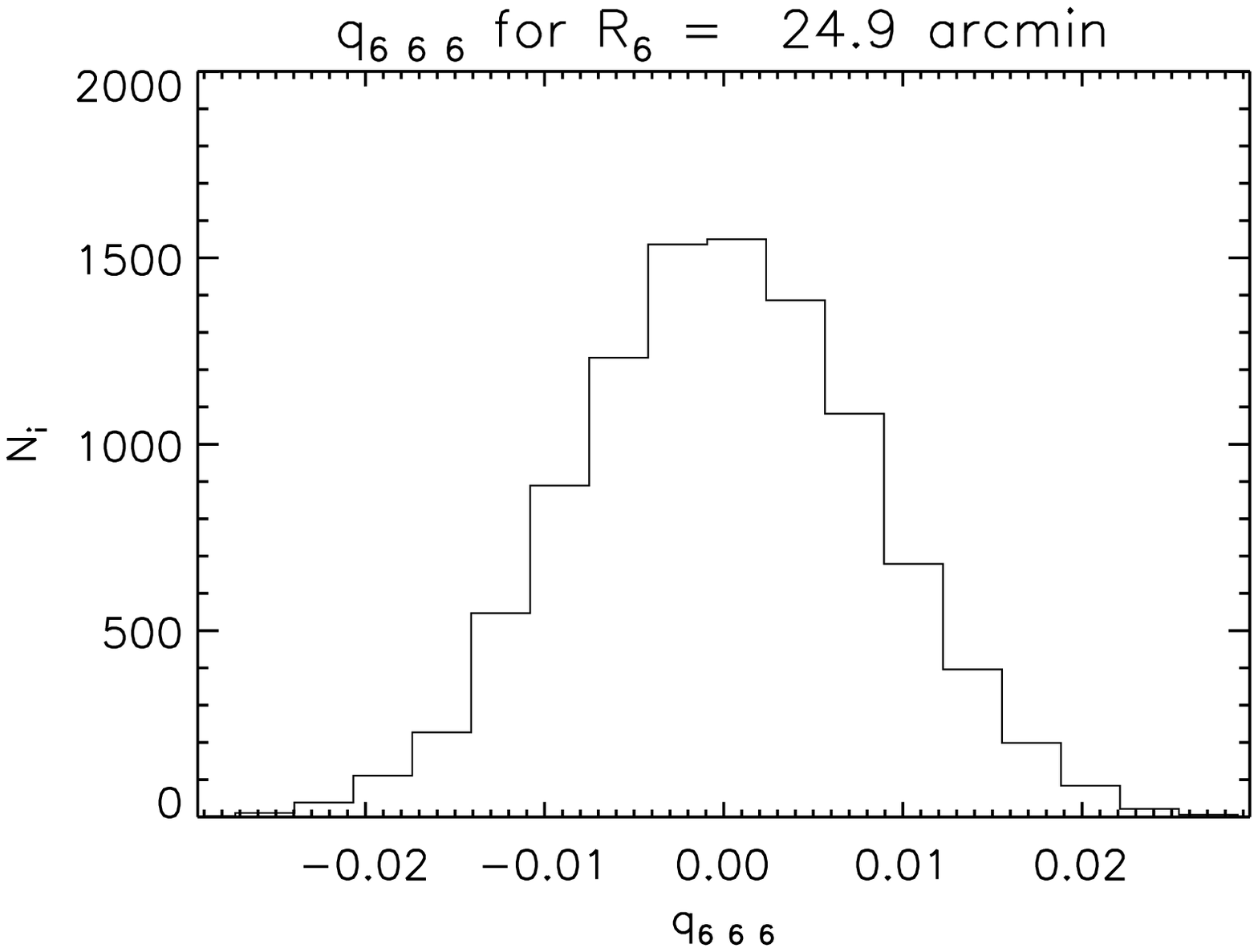}
\includegraphics[width=4.0cm,height=3.0cm, angle = 0] {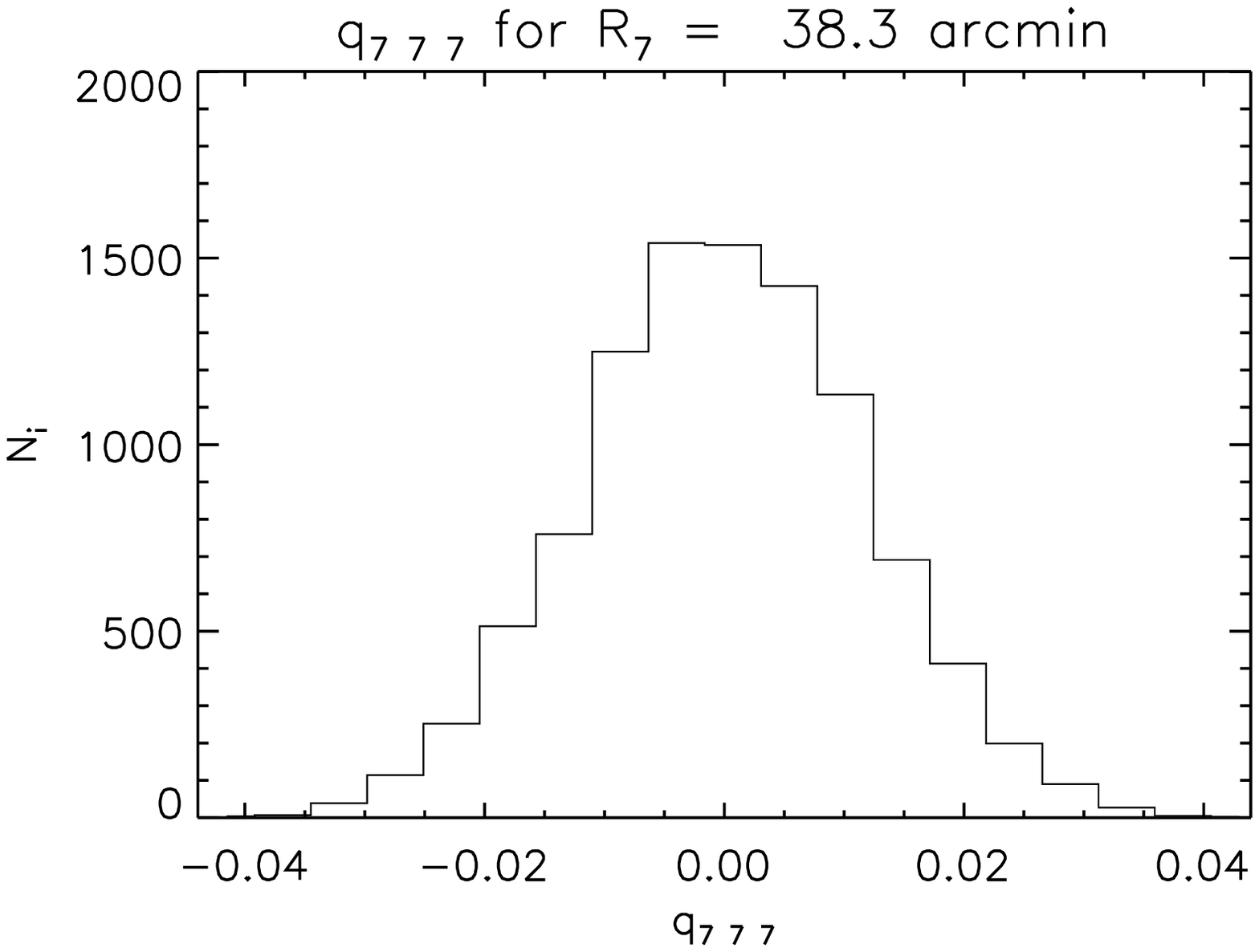}
\includegraphics[width=4.0cm,height=3.0cm, angle = 0] {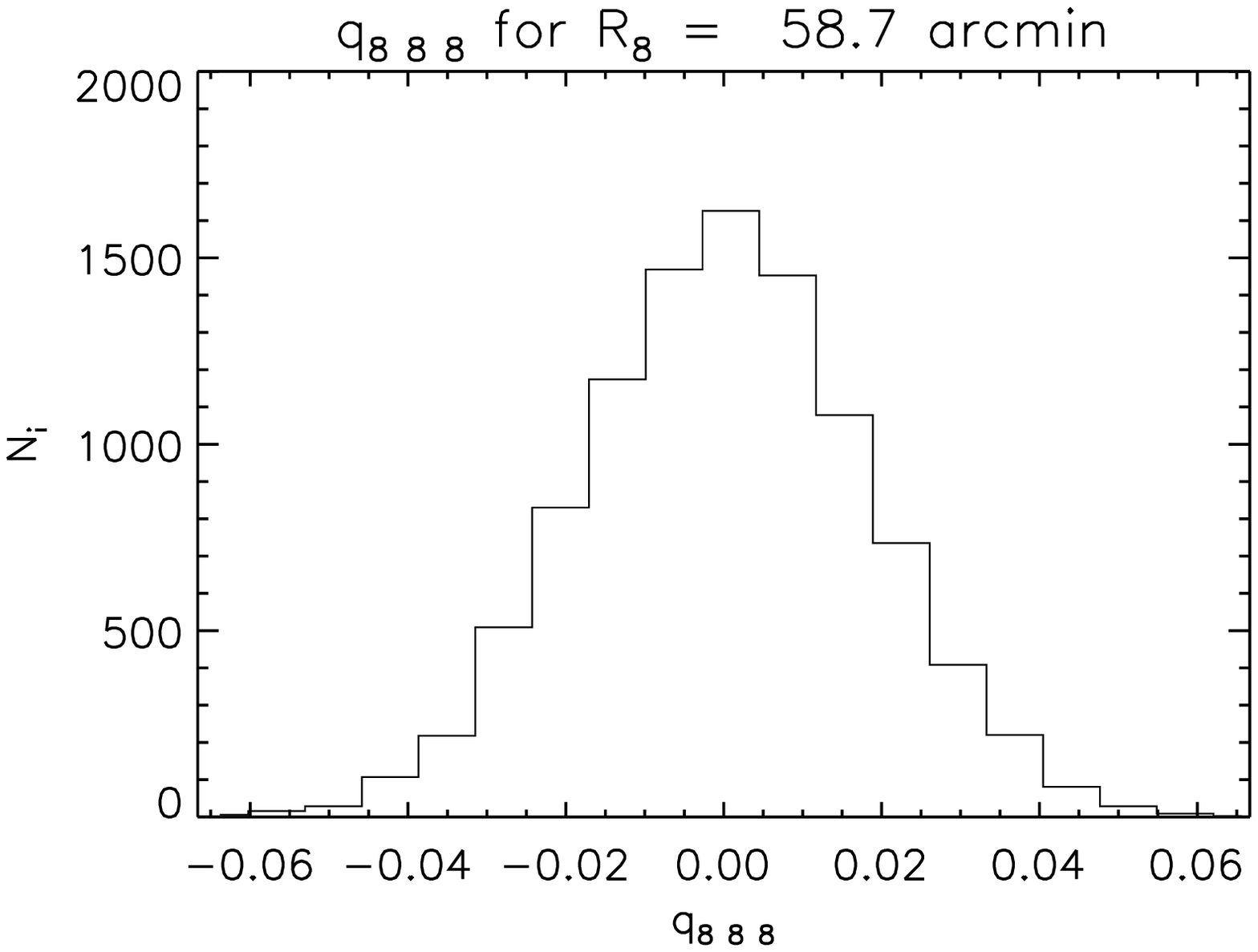}
\includegraphics[width=4.0cm,height=3.0cm, angle = 0] {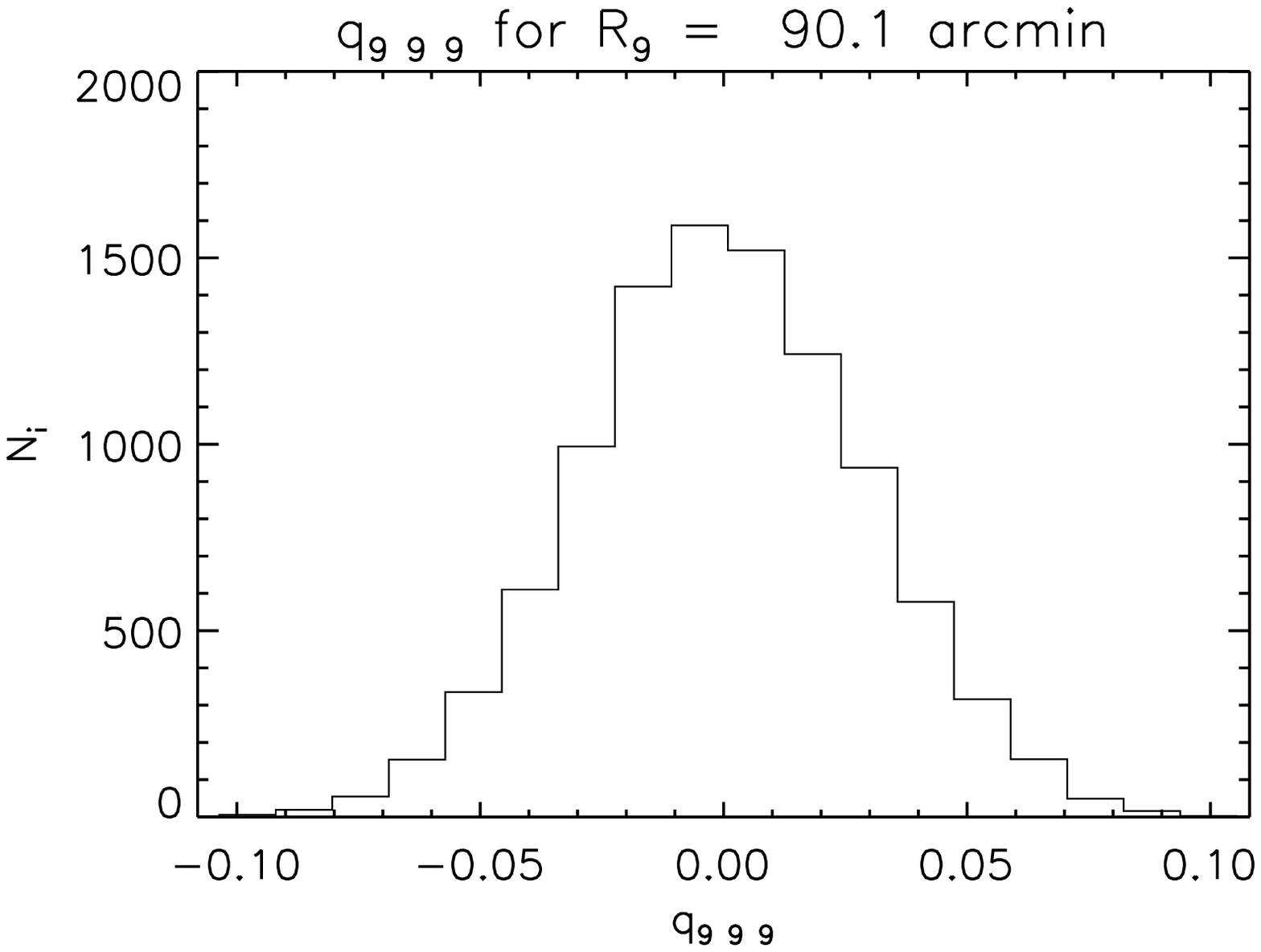}
\includegraphics[width=4.0cm,height=3.0cm, angle = 0] {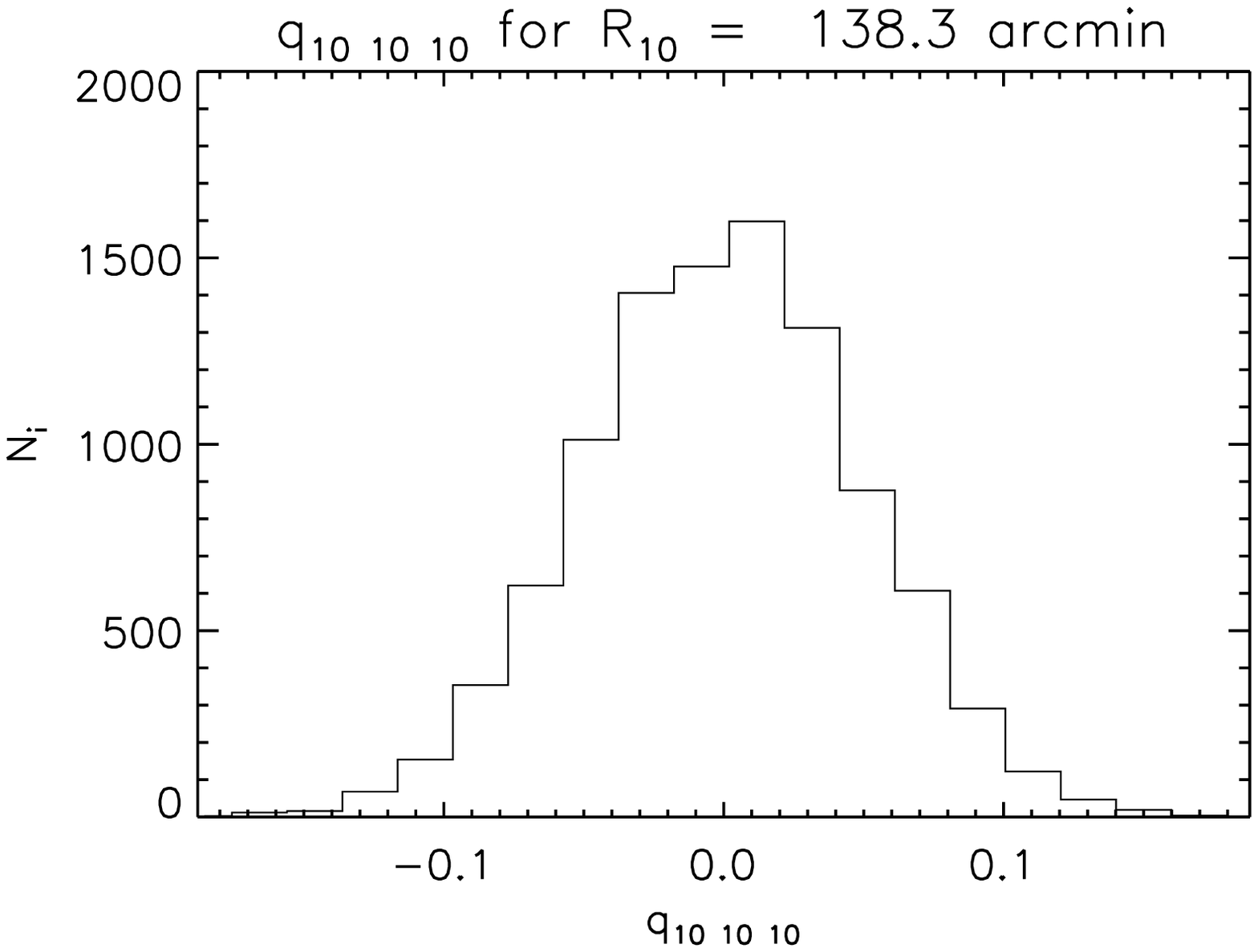}
\includegraphics[width=4.0cm,height=3.0cm, angle = 0] {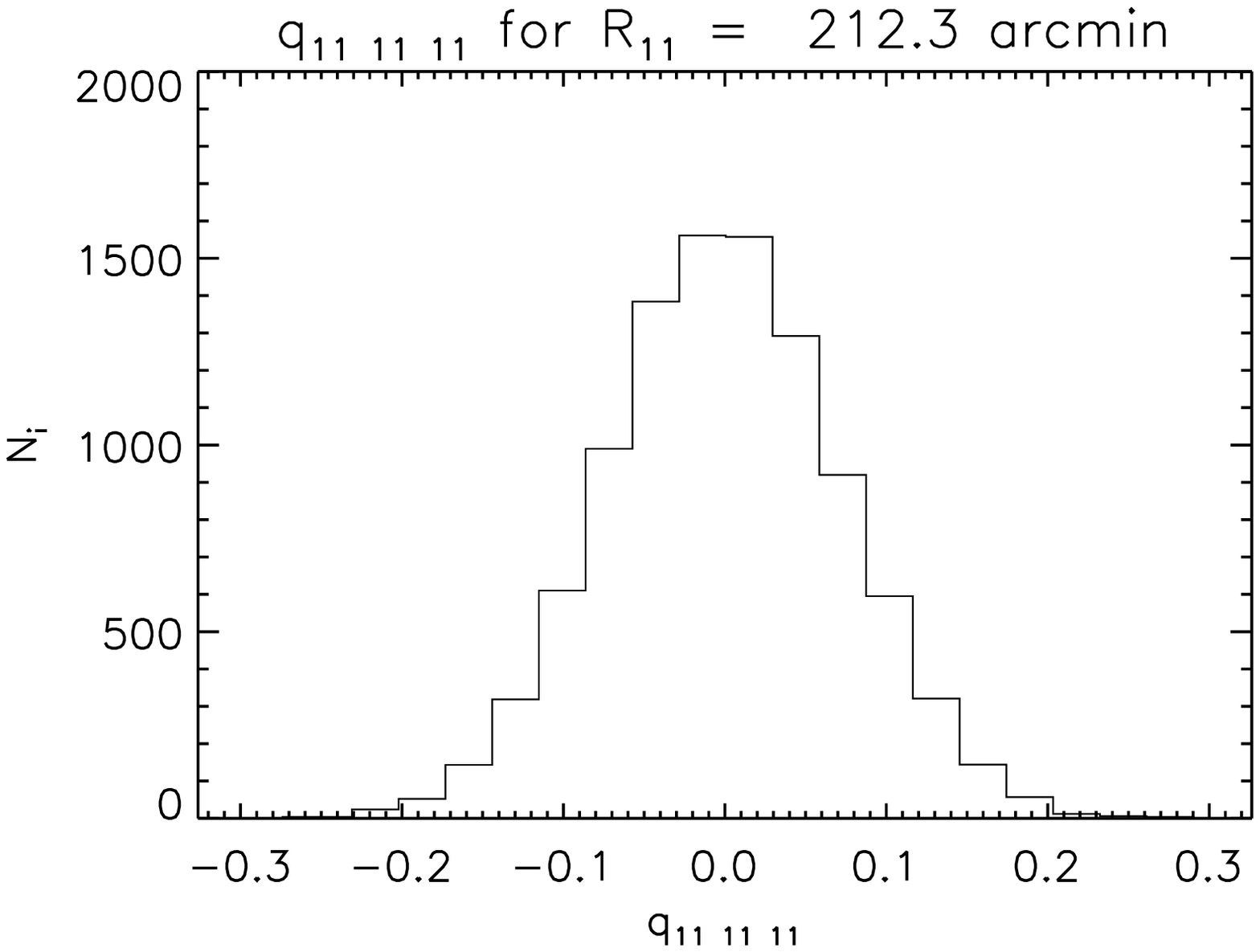}
\includegraphics[width=4.0cm,height=3.0cm, angle = 0] {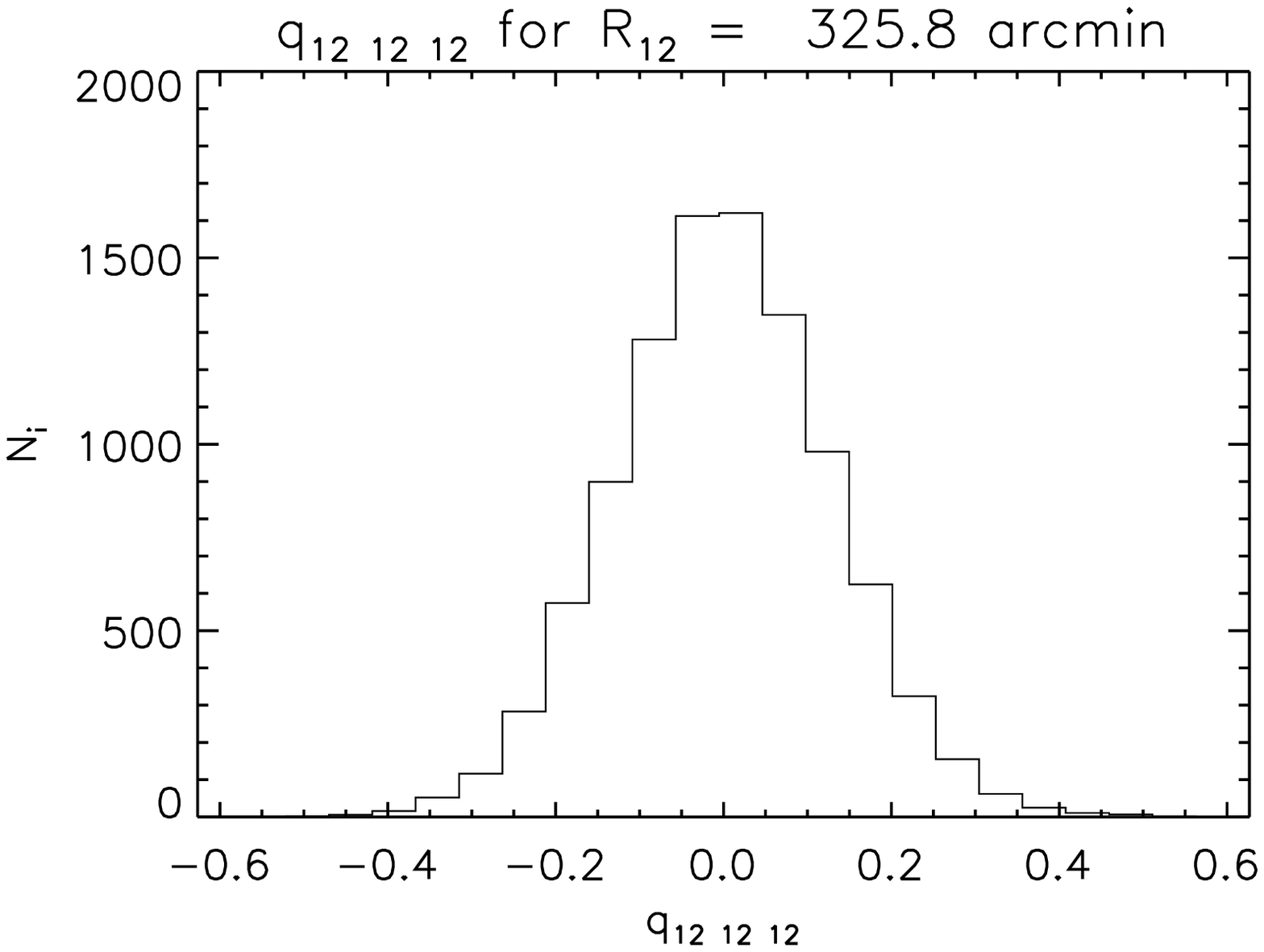}
\includegraphics[width=4.0cm,height=3.0cm, angle = 0] {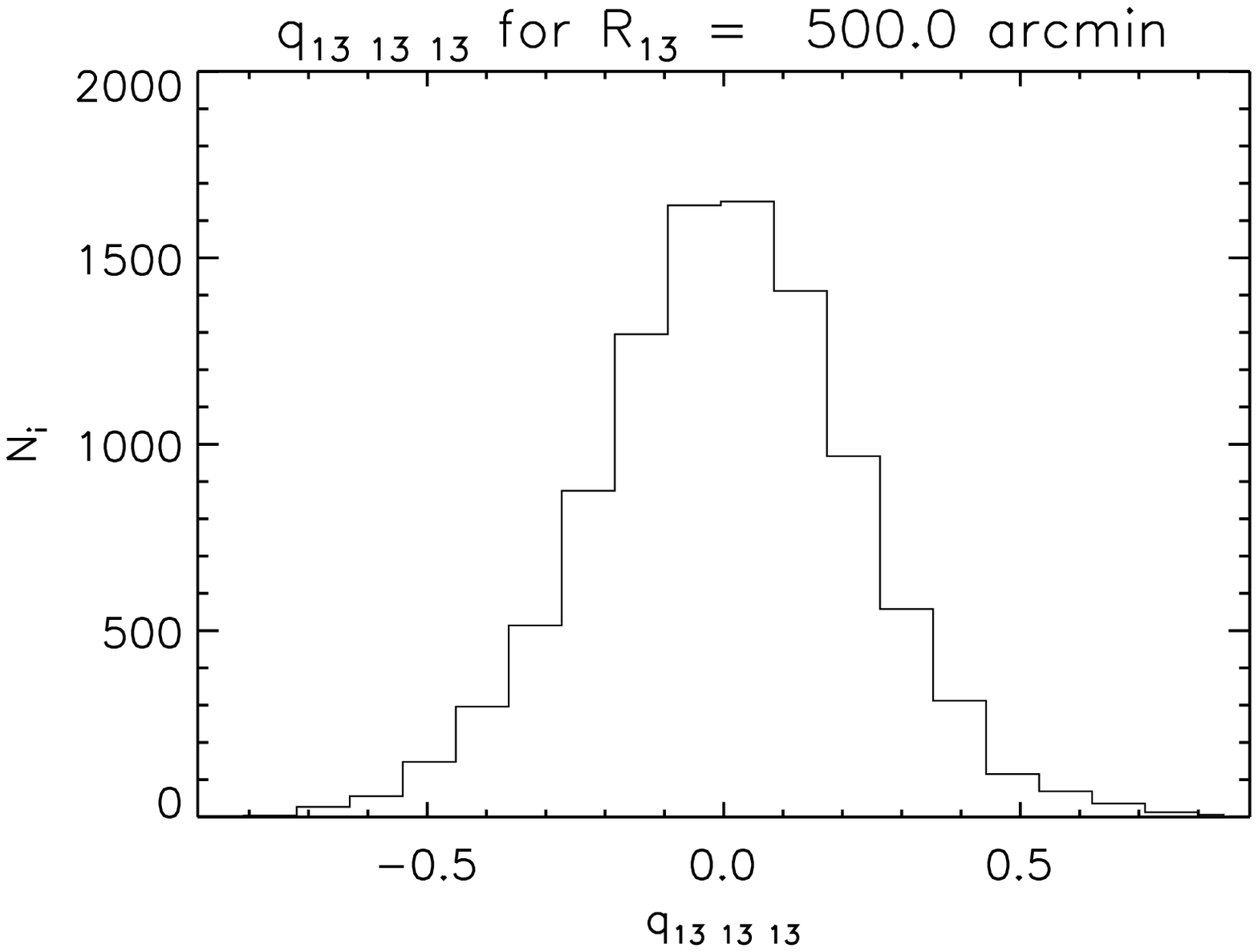}
\includegraphics[width=4.0cm,height=3.0cm, angle = 0] {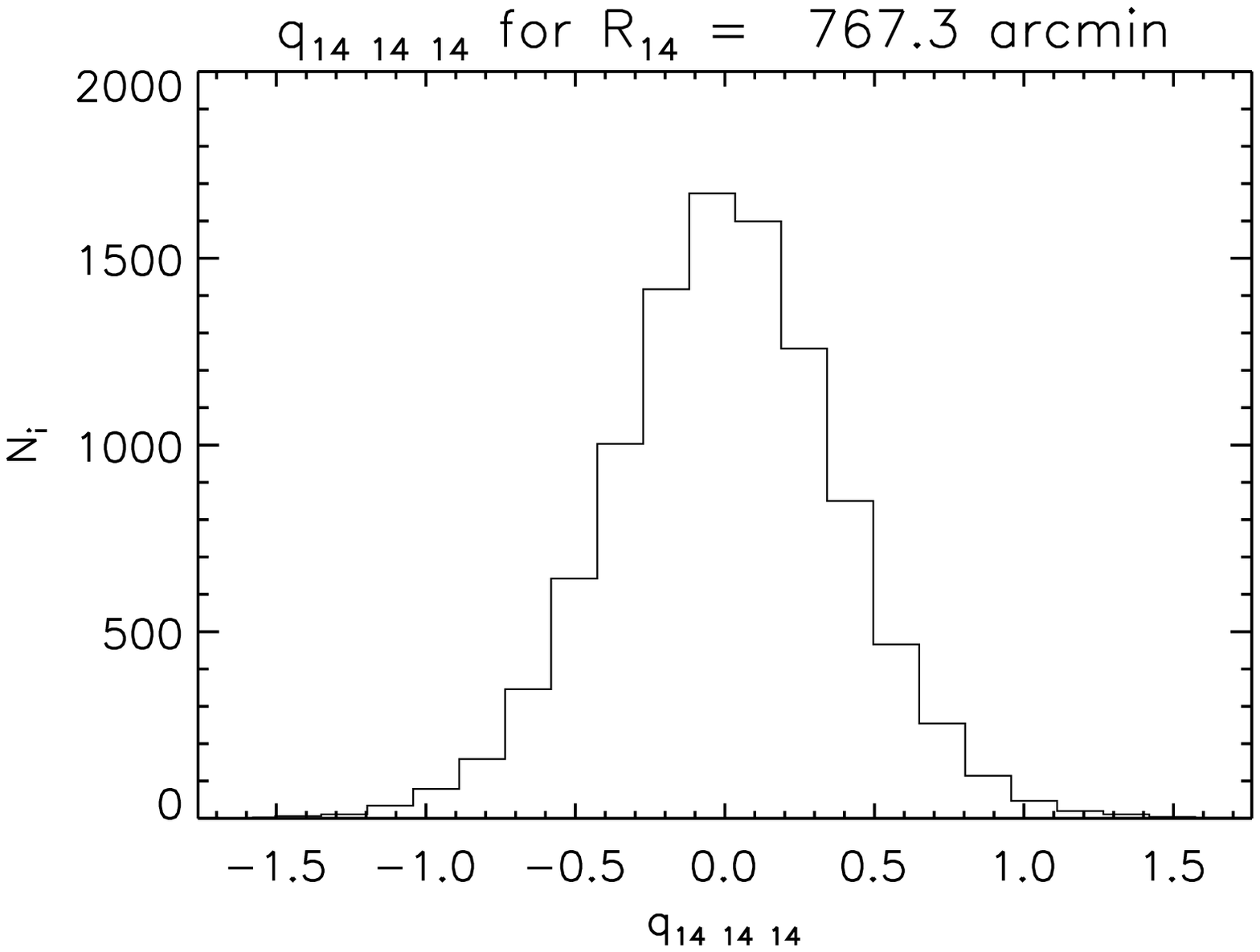}
\caption{Histograms of the cubic statistics $q_{iii}$ for the different angular scales $R_i$ considered in previous works \citep{curto2011a,curto2011b}.
\label{histo_qiii}}
\end{center}
\end{figure*}

Considering the strong decorrelation produced by the convolution of
the SMHW on the temperature anisotropies, we can now apply the central
limit theorem to the cubic statistics defined in
Eq. (\ref{themoments_qijk}). Since the average value is calculated
from the sum of a very large number of almost independent elements (of
the order of the number of pixels in the sphere with size that of the
resolution scale $R$), then its distribution should be very close to a
Gaussian. This is actually seen in Fig. \ref{histo_qiii}, where the
distribution of the cubic terms for different SMHW scales are
shown. These distributions have been obtained from MC simulations of
Gaussian temperature anisotropies.

The previous results indicate that, for Gaussian temperature
anisotropies, a good representation of the n-point distribution of the
quantities $q_{ijk}$ can be given in terms of a multinormal
distribution. Allowing now for the presence of weak non-Gaussianity
for the temperature anisotropies (e.g. an amplitude for the primordial
non-Gaussianity consistent with WMAP data) one can use the next
likelihood for the $f_{nl}$ parameter:
\begin{equation}
L(f_{nl}) \propto e^{-\chi^2(f_{nl})/2},
\label{likelihood}
\end{equation}
where $\chi^2(f_{nl})$ is given by
\begin{equation} 
\chi^2(f_{nl}) = \sum_{ijk,rst}(q_{ijk}^{obs}-f_{nl} \alpha_{ijk})C^{-1}_{ijk, rst}(q_{rst}^{obs}-f_{nl} \alpha_{rst}),
\label{chi_analytic}
\end{equation}
where $q_{ijk}^{obs}$ are the cubic statistics corresponding to the
observed data, $\alpha_{ijk} = \langle q_{ijk} \rangle_{f_{nl}=1}$ and
$C_{ijk, rst}$ is the covariance matrix of the cubic statistics. A
further test to check that the $q_{ijk}$ are normally distributed can
be done by considering the property that $\Delta \chi^2(f_{nl}) =
\chi^2(f_{nl}) - \chi_{min}^2(f_{nl})$ is a $\chi^2$ distribution with
one degree of freedom. In particular, $\Delta \chi^2(f_{nl}) = 1(4)$
should provide the 1(2)-sigma or 68 per cent(95 per cent) confidence
intervals for the $f_{nl}$ parameter. Using MC simulations we have
checked that this is the case for the $q_{ijk}$ statistics.

After straightforward calculation, it can be easily seen that the
$f_{nl}$ estimator in this case is given by
\begin{equation}
\hat{f}_{nl} =
\frac{\sum{\alpha_{ijk}C_{ijk,rst}^{-1}q_{rst}}}{\sum{\alpha_{ijk}C_{ijk,rst}^{-1}\alpha_{rst}}}
\label{estimator_nolinearterm}
\end{equation}
while that the variance of the $\hat{f}_{nl}$ parameter in
Eq. (\ref{estimator_nolinearterm}) is given by
\begin{eqnarray}
\nonumber
\sigma_F^2(\hat{f}_{nl}) = \frac{-1}{\Big\{\frac{\partial^2 log L(f_{nl})}{\partial
    f_{nl}^2}\Big\}_{f_{nl}=\hat{f}_{nl}}}  =  \\
\frac{1}{\frac{1}{2}{\Big\{\frac{\partial^2
   \chi^2(f_{nl})}{\partial f_{nl}^2}\Big\}}_{f_{nl}=\hat{f}_{nl}}} = \frac{1}{\sum_{ijk,rst}\alpha_{ijk}C^{-1}_{ijk,rst}\alpha_{rst}}.
\label{sigma_fnl_likelihood}
\end{eqnarray}
This estimator has already been shown to be nearly optimal on WMAP
data \citep{curto2009a,curto2009b,curto2010,curto2011a,curto2011b}
without the need of subtracting any linear term. However, as stated in
\citet{donzelli2012}, from all the possible cubic combinations of
three Gaussian variables, the Wick polynomials are shown to have
minimum variance. This implies that in order to have a strictly
speaking minimum variance estimator, a linear term correction needs to
be included. In fact the linear term subtraction is equivalent to the
mean subtraction at each wavelet coefficient map \citep{donzelli2012}
for low levels of anisotropy. This is indeed the procedure that has
been followed in
\citet{curto2009a,curto2009b,curto2010,curto2011a,curto2011b} and it
explains the competitive results obtained just by subtracting the mean
using the estimator in Eq. \ref{estimator_nolinearterm}.

The linear term correction for the wavelet estimator can be written as:
\begin{equation}
 \hat{f}_{nl}^{(total)} =  \hat{f}_{nl}^{(cubic)} - \hat{f}_{nl}^{(linear)},
\label{estimator_linear_term}
\end{equation}
where $ \hat{f}_{nl}^{(cubic)}$ is given by
Eq. \ref{estimator_nolinearterm} and
\begin{equation}
 \hat{f}_{nl}^{(linear)} =  \sigma^2_{F}(\hat{f}_{nl}) \sum_{ijk, rst} \alpha_{ijk} C_{ijk, rst}^{-1} q_{rst}^{(L)}
\end{equation}
with
\begin{eqnarray}
\nonumber
q_{ijk}^{(L)} = \frac{1}{N_{ijk}}\sum_{p}  \bigg \{ \Big \langle \frac{w_i(p)}{\sigma_i}\frac{w_j(p)}{\sigma_j} \Big \rangle \frac{w_k(p)}{\sigma_k} + \\
\Big \langle \frac{w_i(p)}{\sigma_i}\frac{w_k(p)}{\sigma_k} \Big \rangle \frac{w_j(p)}{\sigma_j} + \Big \langle \frac{w_j(p)}{\sigma_j}\frac{w_k(p)}{\sigma_k} \Big \rangle \frac{w_i(p)}{\sigma_i}\bigg \}.
\end{eqnarray}
In the next Sections we apply the wavelet estimator to WMAP 7-year
data as well as to Planck simulations and compare the results
obtained with and without the linear term correction.
\section{Application to WMAP V+W data}
\label{sec:wmap_data}
We have computed the linear term correction to the cubic wavelet
$f_{nl}$ estimator for the three shapes with a relevant interest in
many inflationary models: the local, equilateral and orthogonal shapes
\citep[see for example][]{bartolo2004,senatore2010,komatsu2011}. The
estimator can be easily applied to other bispectra with a separable
shape \citep{curto2011a}. Results taking into account only the cubic
contribution are already published \citep{curto2011b}.

We have selected the same set of 15 angular scales from $R_0 = 0$
arcmin to $R_{14} = 767.3$ arcmin used in \citet{curto2011b}. We have
considered the V+W WMAP data optimally weighted by the $N_{hits}$ maps
per radiometer in order to maximise the signal-to-noise ratio. We also
consider the same WMAP $KQ75$ mask and its extended masks for each
wavelet angular scale. The cubic covariance matrix has been computed
using 10000 Gaussian simulations. A principal component analysis has
been performed in order to avoid contamination from the lowest noisy
eigenvalues of this covariance matrix without losing non-Gaussian
signal \citep{curto2011a}. The two point correlation matrices needed
for the linear term correction have also been estimated with 64000
Gaussian simulations. This number of simulations is needed in order to
achieve the required precision in the estimation of the correlation
matrices.

\begin{figure*}
\centering
\includegraphics[width=7.0cm, height=4.4cm, angle= 0]{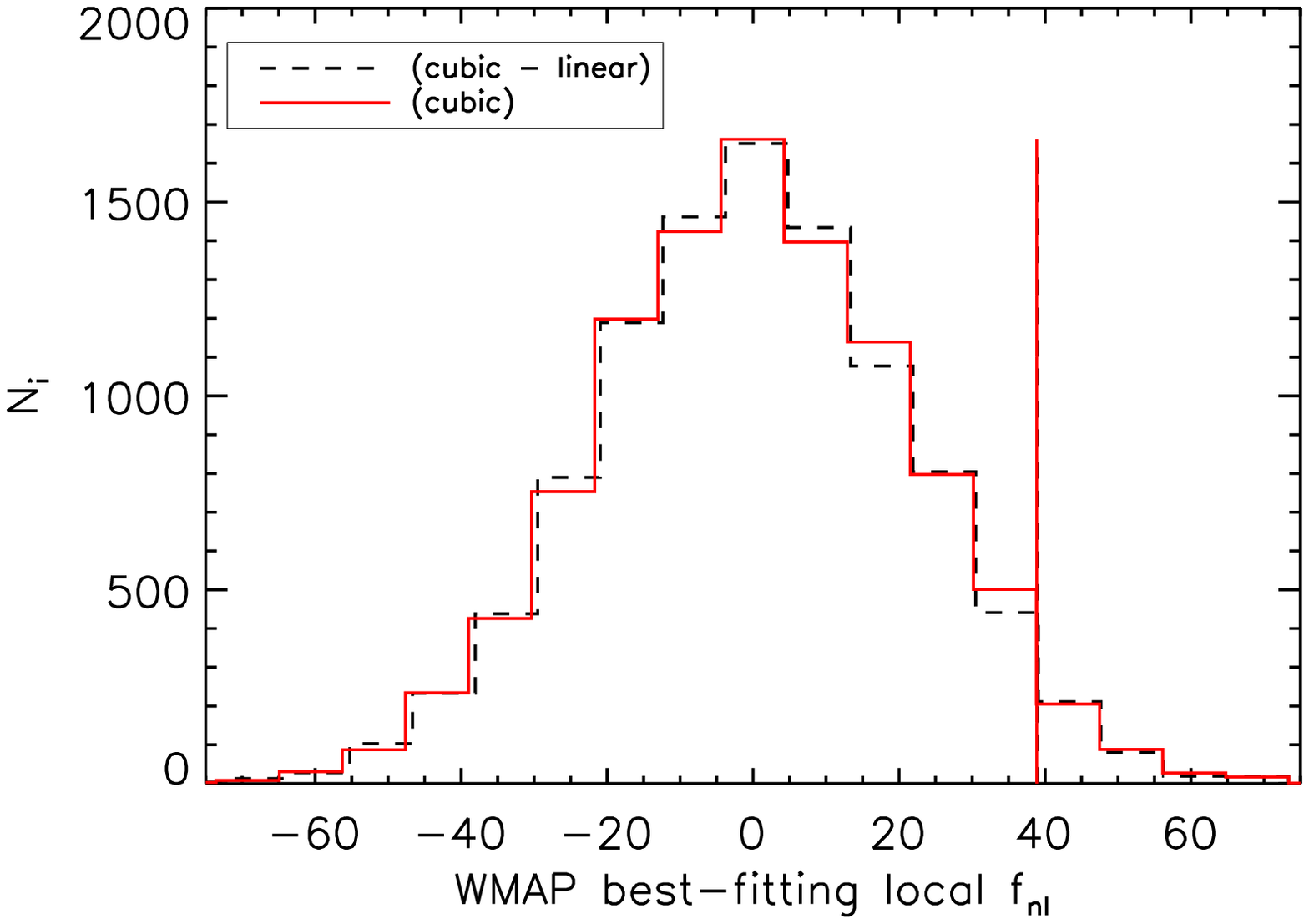}
\includegraphics[width=7.0cm, height=4.4cm, angle= 0]{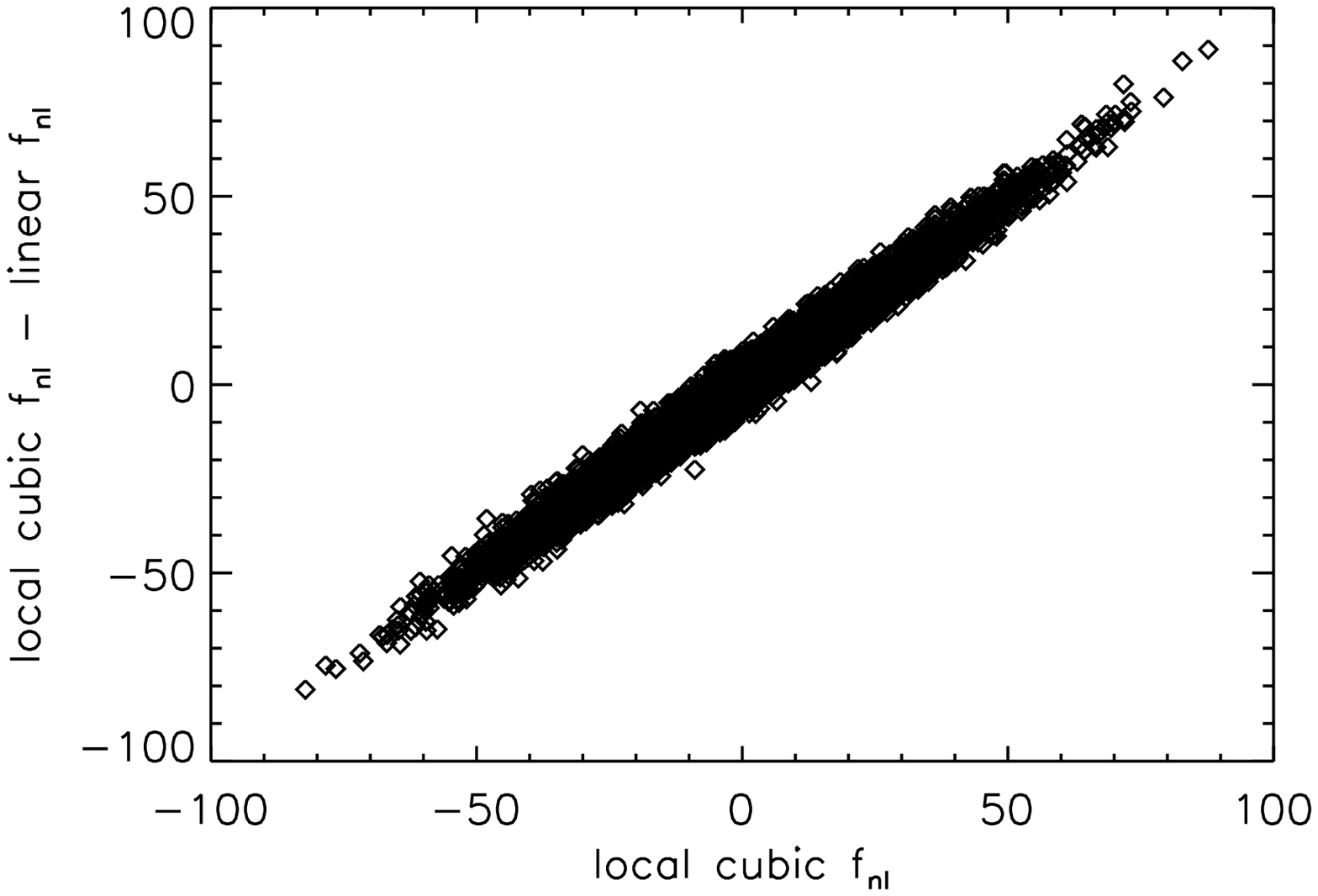}
\includegraphics[width=7.0cm, height=4.4cm, angle= 0]{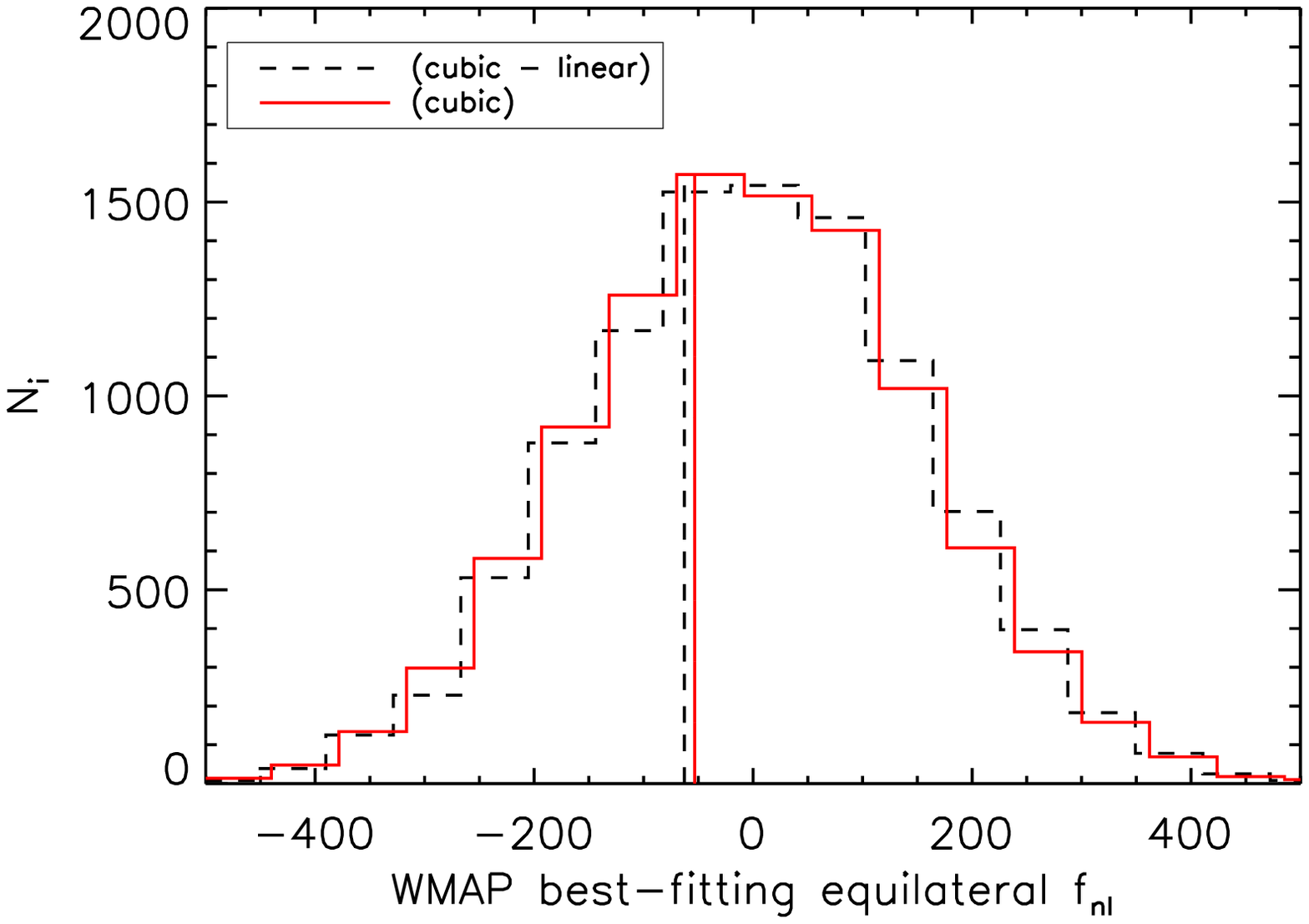}
\includegraphics[width=7.0cm, height=4.4cm, angle= 0]{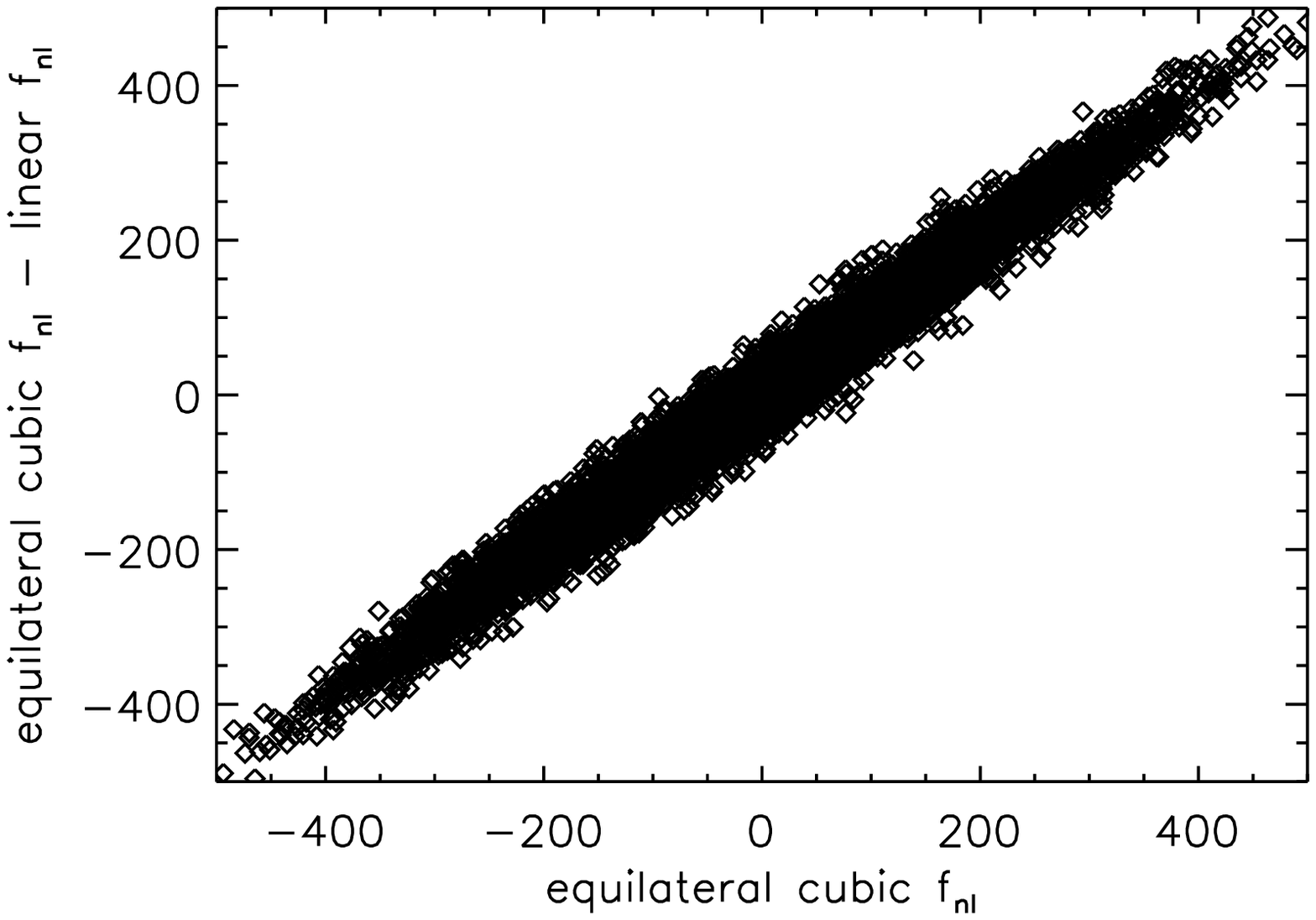}
\includegraphics[width=7.0cm, height=4.4cm, angle= 0]{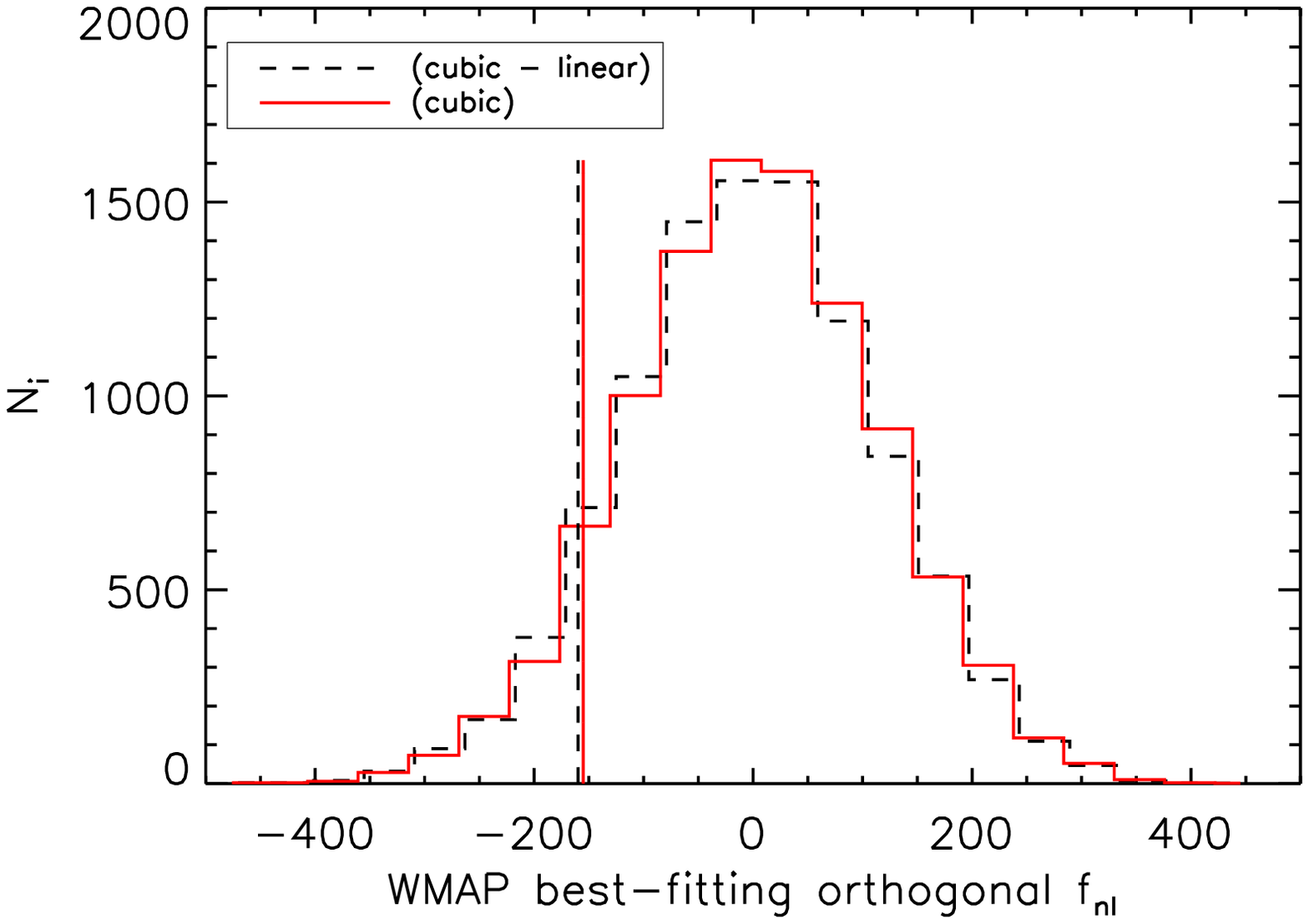}
\includegraphics[width=7.0cm, height=4.4cm, angle= 0]{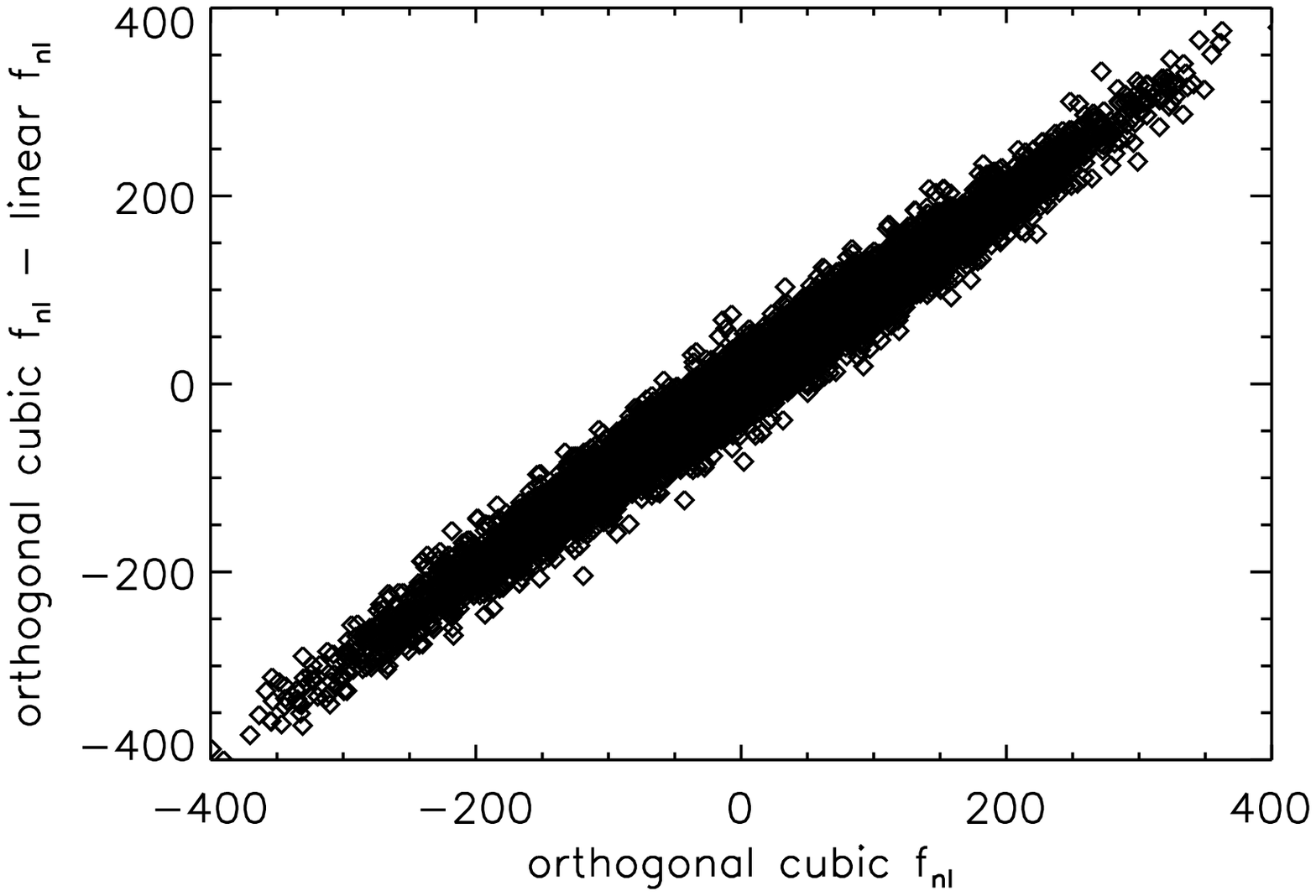}
\caption{WMAP 7-year data best-fitting $f_{nl}$ values using the cubic
  estimator (Eq. \ref{estimator_nolinearterm}) and the estimator with
  the linear correction (Eq. \ref{estimator_linear_term}) for the
  local ({\it top}), equilateral ({\it middle}) and orthogonal ({\it
    bottom}) shapes. In the left panels the histogram of the
  best-fitting $f_{nl}$ values with (dashed dark line) and
  without (solid red line) the linear term correction for each
  simulation are plotted. Vertical lines correspond to the values
    obtained with WMAP data. The right pannels show the corresponding
    correlation between the same estimates.}
\label{fig:cubic_linear_fnl}       
\end{figure*}

We have applied the estimator to one set of 10000 Gaussian maps and
the WMAP data. The results are presented in
Fig. \ref{fig:cubic_linear_fnl} for the three considered shapes. In
the left panels, the red histograms correspond to the best-fitting
$f_{nl}$ values obtained with the cubic estimator and the black
histograms correspond to the best-fitting $f_{nl}$ values after the
linear term correction. The vertical lines correspond to the actual
WMAP data values estimated with the cubic estimator (red) and the
linearly corrected estimator (black). In the right panels, we compare
the best-fitting $f_{nl}$ values for the same set of Gaussian
simulations. Note that both $f^{(cubic)}_{nl}$ and $f^{(total)}_{nl}$
are highly correlated and the deviations are not significant.
Finally in Tables \ref{table_fnl_results_local},
\ref{table_fnl_results_equilat} and \ref{table_fnl_results_ortho} the
previous results are summarised. We present the WMAP 7-year $f_{nl}$
best-fitting values for the cubic estimator $f^{(cubic)}_{nl}$, the
linear estimator $f^{(linear)}_{nl}$, and $f^{(total)}_{nl}$ for the
clean and raw (uncleaned) maps. The Fisher $f_{nl}$ error bar as
described in Eq. 25 of \citet{curto2011a} is also provided.  For each
case, we observe a small reduction of the error bars when the linear
term is included. The largest correction is introduced in the local
shape, where $\sigma$ is reduced from $\sigma(f_{nl}) = 21.6$ to
$\sigma(f_{nl}) = 21.4$ \citep[i.e. a reduction of 1 per cent, in
  agreement with][]{donzelli2012}. The correction for the other two
cases, equilateral and orthogonal is also negligible (about 0.2 per
cent and 0.1 per cent respectively). This is in agreement with
\citet{creminelli2006} for the equilateral shape, where the standard
deviations of $f_{nl}$ without the linear term were found closer to
the lower Fisher limit than in the local shape, suggesting a less
important contribution of the linear term correction.

Our best-fitting values, computing $\sigma(f_{nl})$ with 10000
Gaussian simulations to characterize the errors, are presented below
for the three shapes. 

{\bf Local form results}\footnote{The average $\langle q_{ijk}
  \rangle_{f_{nl}}$ is obtained using 1000 non-Gaussian simulations of
  the local shape generated by the procedure described in
  \citep{elsner2009} and publicly available at
  http://planck.mpa-garching.mpg.de/cmb/fnl-simulations/ 

The best estimates of the local shape presented in \citet{curto2011b}
are $f^{(cubic)}_{nl} = 32.5 \pm 22.5$. Note that in that work, a
perturbative approach is considered to simulate the non-Gaussian
simulations used to compute $\langle q_{ijk} \rangle_{f_{nl}}$. The
different approaches to simulate the non-Gaussianity and the
statistical errors due to the finite number of non-Gaussian
simulations explains the small differences between the error bars
presented here and in that reference.}:
\begin{itemize}
\item $ f^{(cubic)}_{nl} = 38.9 \pm 21.6$
\item $ f^{(total)}_{nl} = 39.0 \pm 21.4$
\end{itemize}

{\bf Equilateral form results:}
\begin{itemize}
\item $ f^{(cubic)}_{nl} = -53.3 \pm 154.3$
\item $ f^{(total)}_{nl} = -62.8 \pm 154.0$
\end{itemize}

\begin{figure*}
\centering
\includegraphics[width=5.5cm, height=4.4cm, angle= 0]{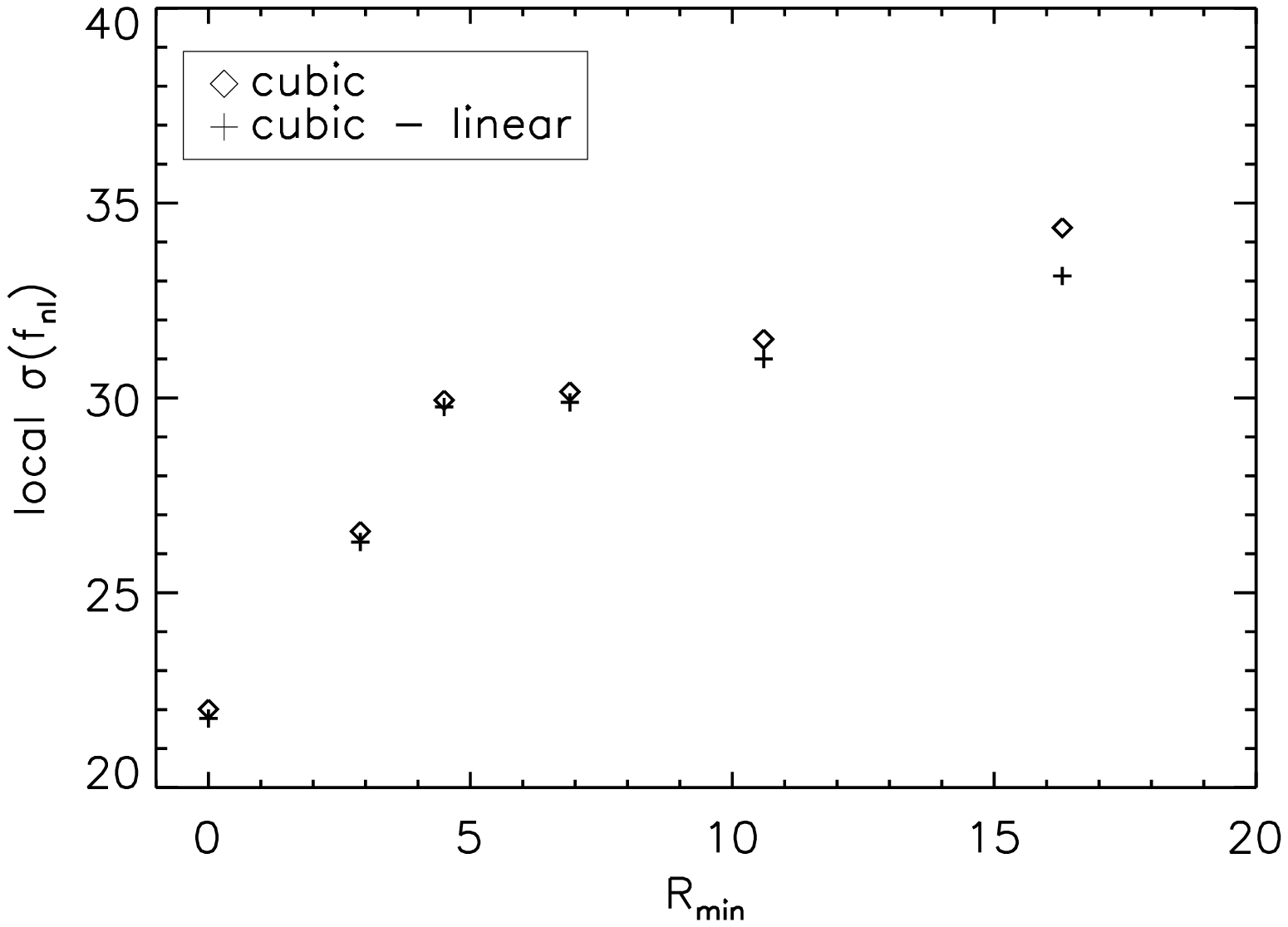}
\includegraphics[width=5.5cm, height=4.4cm, angle= 0]{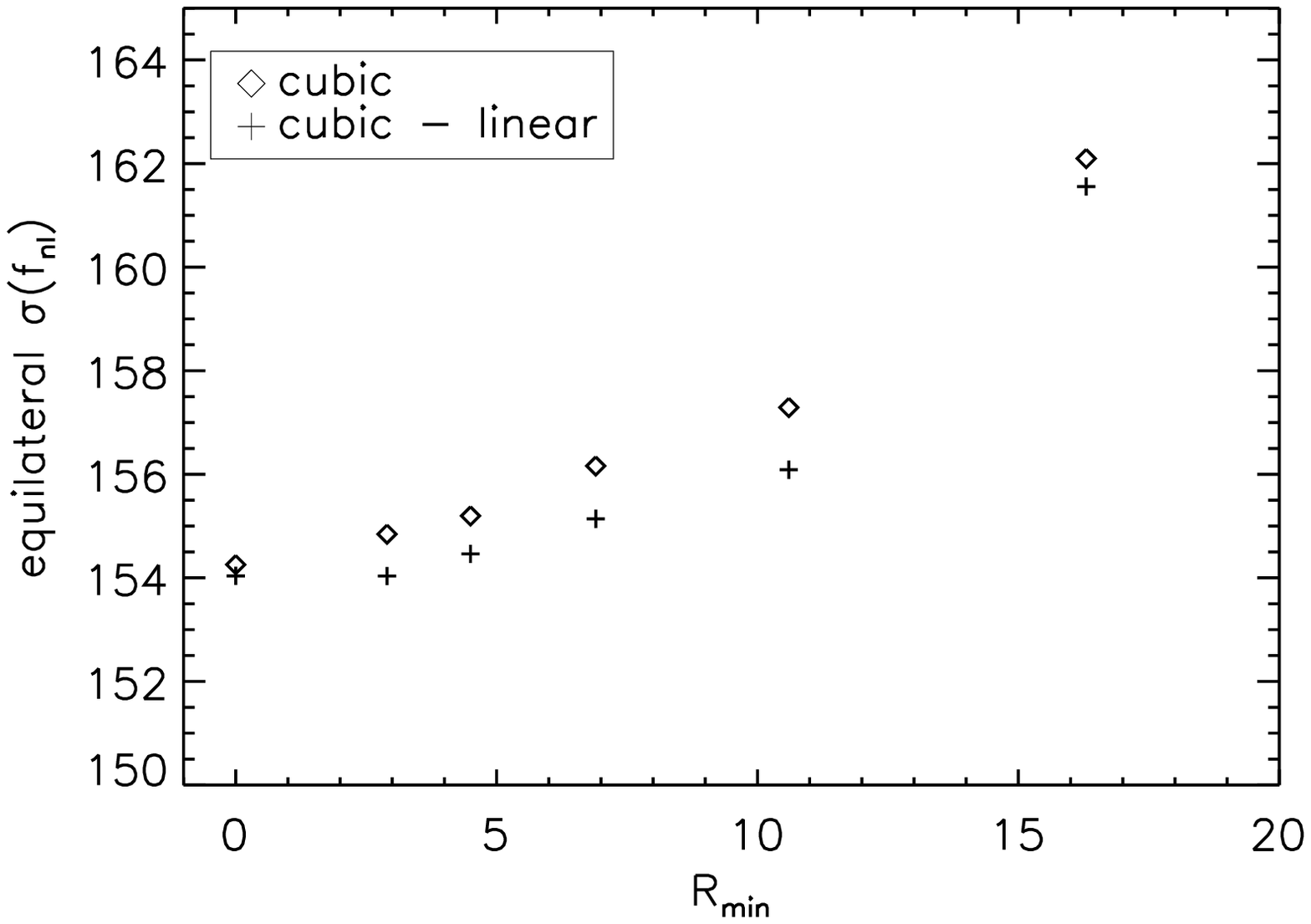}
\includegraphics[width=5.5cm, height=4.4cm, angle= 0]{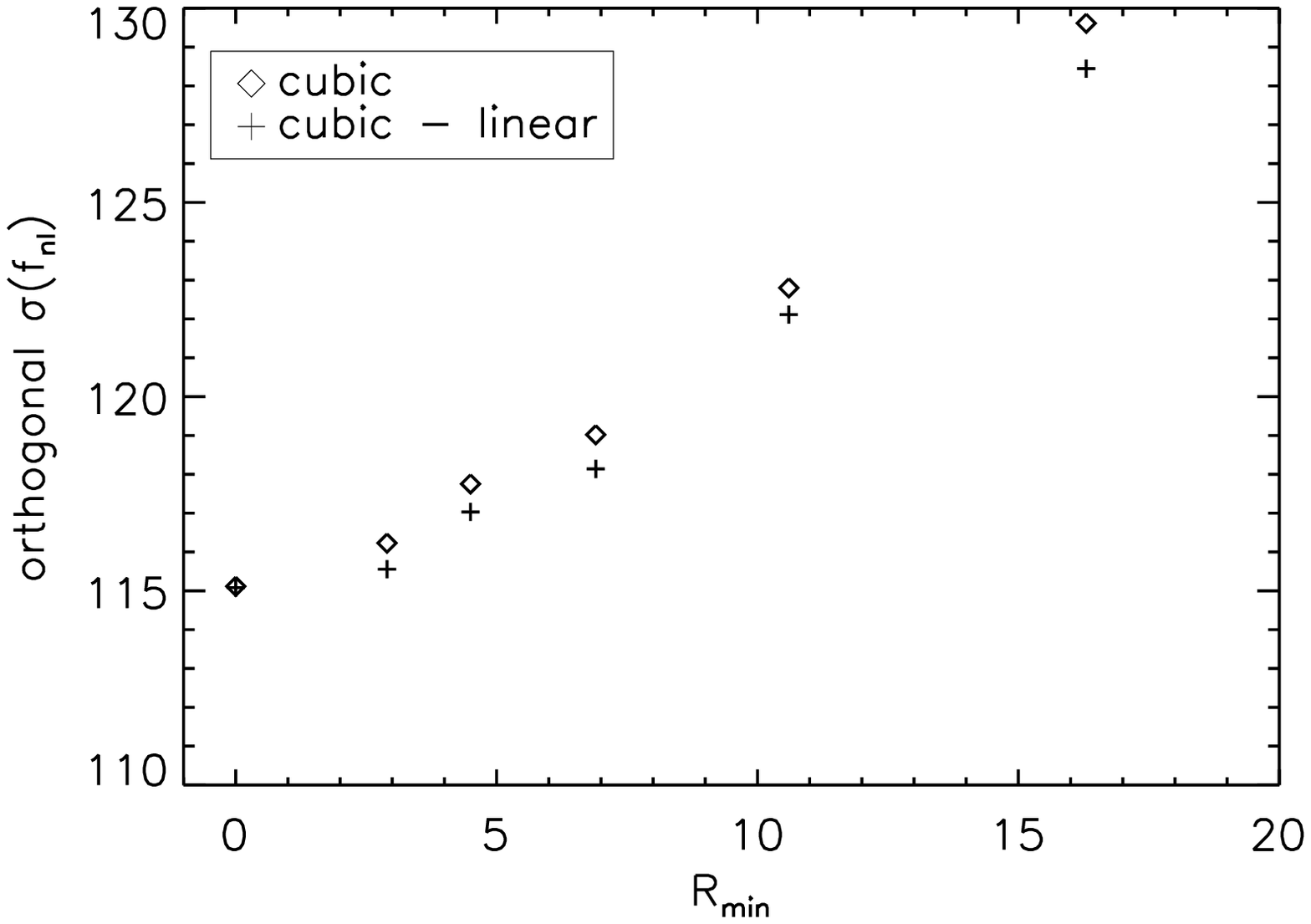}
\caption{{The $f_{nl}$ error bars computed using different $R_{min}$
  angular scales for the local {\it (left)}, equilateral {\it (middle)}
  and orthogonal {\it (right)} shapes. The diamonds corresponds to
  $f^{cubic}_{nl}$ and the crosses to $f^{total}_{nl}$.}}
\label{fig:sig_vs_Rmin}       
\end{figure*}

{\bf Orthogonal form results:}
\begin{itemize}
\item $ f^{(cubic)}_{nl} = -155.1 \pm 115.1$
\item $ f^{(total)}_{nl} = -159.8 \pm 115.1$
\end{itemize}
\begin{table*}
  \center
  \caption{Constraints on the $f_{nl}$ parameter for the local shape
    with and without the linear term correction. From left to
      right, the best-fitting values for the clean and the raw data
      maps, the mean, dispersion, 16, 84, 2.5 and 97.5 per cent quantiles
      respectively of the $f_{nl}$ distribution obtained with 10000
      Gaussian maps. The Fisher error bar obtained for this shape is
      $\sigma_F(f_{nl})=21.6$.
\label{table_fnl_results_local}}
\begin{tabular}{c|cc|cccccc}
\hline
case & $f_{nl}^{(clean~data)}$ & $f_{nl}^{(raw~data)}$& $\langle f_{nl} \rangle$ & $\sigma(f_{nl})$  & $X_{16}$ & $X_{84}$ & $X_{2.5}$ & $X_{97.5}$ \\
\hline
cubic & 38.9& 20.8 & 0.6 & 21.6 & -21.1 & 21.9 & -42.7 & 41.8 \\
linear & -0.1 & -0.0  &  0.0  & 3.1 & -3.1 & 3.2 & -6.1 & 6.2 \\
cubic - linear & 39.0 & 20.8 & 0.7 & 21.4 & -21.0 & 22.7 & -42.6 & 41.3 \\
\hline
\end{tabular}
  \caption{Constraints on the $f_{nl}$ parameter for the equilateral
    shape with and without the linear term correction. From left to
    right, the best-fitting values for the clean and the raw data
    maps, the mean, dispersion, 16, 84, 2.5 and 97.5 per cent
    quantiles respectively of the $f_{nl}$ distribution obtained with
    10000 Gaussian maps. The Fisher error bar obtained for this shape
    is $\sigma_F(f_{nl})=144.5$.
\label{table_fnl_results_equilat}}
\begin{tabular}{c|cc|cccccc}
\hline
case & $f_{nl}^{(clean~data)}$ & $f_{nl}^{(raw~data)}$& $\langle f_{nl} \rangle$ & $\sigma(f_{nl})$ & $X_{16}$ & $X_{84}$ & $X_{2.5}$ & $X_{97.5}$ \\
\hline
cubic & -53.3 & 28.1 & -1.6 & 154.3 & -155.9 & 151.5 & -302.4 & 302.2 \\
linear & 9.5 & 13.7 & -0.3 & 23.0 & -23.6 & 22.6 & -47.8 &  46.7\\
cubic - linear & -62.8 & 14.4 & -1.3 & 154.0 & -156.4 & 150.3 & -304.5 & 300.3 \\
\\
\hline
\end{tabular}
  \caption{Constraints on the $f_{nl}$ parameter for the orthogonal
    shape with and without the linear term correction. From left
      to right, the best-fitting values for the clean and the raw data
      maps, the mean, dispersion, 16, 84, 2.5 and 97.5 per cent quantiles
      respectively of the $f_{nl}$ distribution obtained with 10000
      Gaussian maps. The Fisher error bar obtained for this shape is
      $\sigma_F(f_{nl})=106.3$.
\label{table_fnl_results_ortho}}
\begin{tabular}{c|cc|cccccc}
\hline
case & $f_{nl}^{(clean~data)}$ & $f_{nl}^{(raw~data)}$& $\langle f_{nl} \rangle$ & $\sigma(f_{nl})$ & $X_{16}$ & $X_{84}$ & $X_{2.5}$ & $X_{97.5}$ \\
\hline
cubic & -155.1 & -119.4 & 0.2 & 115.1 & -113.4 & 115.2 & -230.4 & 225.6 \\
linear & 4.7 & 4.8 & 0.2 & 18.5 & -17.8 & 18.0 & -36.5 & 36.7 \\
cubic - linear  & -159.8 & -124.2 & 0.0 & 115.1 & -113.2 & 115.0 & -228.4 & 222.8 \\
\hline
\end{tabular}
\end{table*}
%
In order to check that the estimator has already reached optimality
with the considered scales for the three shapes, we have computed
$\sigma(f_{nl})$ for different subsets of scales
(Fig. \ref{fig:sig_vs_Rmin}). We compare the $f_{nl}$ error bars for
different minimum angular scales $R_{min}$. To find the equivalent
multipole $\ell$ range corresponding to each $R_{min}$ see Fig. 5 of
\citet{curto2011a}. The three shapes reach minimum variance for
$R_{min} = 0$ arcmin.

The error bar of the equilateral and orthogonal shapes are also
similar to the values obtained with the direct bispectrum estimator
where $\sigma(f_{nl}) = 140$ for the equilateral shape and
$\sigma(f_{nl}) = 104$ for the orthogonal shape
\citep{komatsu2011}. The slightly larger values ($\sim 9$ per cent)
obtained from the dispersion of the $f_{nl}$ distribution
corresponding to 10000 Gaussian simulations, $\sigma(f_{nl}) = 154$
and $\sigma(f_{nl}) = 115$ respectively, are likely due to differences
in the perturbative approach used to simulate the non-Gaussian signal
of these two shapes \citep{curto2011b} or the statistical errors due
to the finite number of non-Gaussian simulations.
\section{Application to Planck simulations}
\label{sec:planck_simulations}
We have computed the linear term correction to the cubic wavelet
$f_{nl}$ estimator for the local shape using Planck simulations in
order to forecast the amplitude of this correction on future Planck
analyses. We do not consider the two other shapes (equilateral and
orthogonal). From the results of previous sections, we expect the
correction for those cases to be even smaller.

For this analysis we have considered a new set of angular scales that
better suits the range of angular multipoles which are cosmic variance
dominated ($\ell_{\max} \sim 2000$). The list of angular scales is
$R_0=0$, $R_1=1.3$, $R_2=2.1$, $R_3=3.4$, $R_4=5.4$, $R_5=8.7$,
$R_6=13.9$, $R_7=22.3$, $R_8=35.6$, $R_9=57.0$, $R_{10}=91.2$,
$R_{11}=146.0$, $R_{12}=233.5$, $R_{13}=373.6$, $R_{14}=597.7$ and
$R_{15}=956.3$ arcmin. As a representative mask, we have used the
available WMAP $KQ75$ mask (75 per cent of the sky). We have simulated
the Planck 143 GHz channel using a fiducial CMB power spectrum that
best fits WMAP 7-year data, $\ell_{max} = 2048$ and a Gaussian beam
with $FWHM = 7.1$ arcmin. The noise has been generated using an
anisotropic $N_{hits}$ map computed from the scanning strategy of the
Planck Sky
Model\footnote{http://www.apc.univ-paris7.fr/$\sim$delabrou/PSM/psm.html}
\citep{delabrouille2012} and the noise sensitivity per pixel provided
in the Planck Bluebook\footnote{ The Planck Bluebook is available for
  download in the web:
  http://www.rssd.esa.int/index.php?project=Planck} (using an average
noise sensitivity for 14 months of $\sigma_{noise} = 2.2 ~\mu$K/K in a square pixel whose size is the FWHM extent of the
beam).

\begin{figure*}
\centering
\includegraphics[width=7.0cm, height=4.4cm, angle= 0]{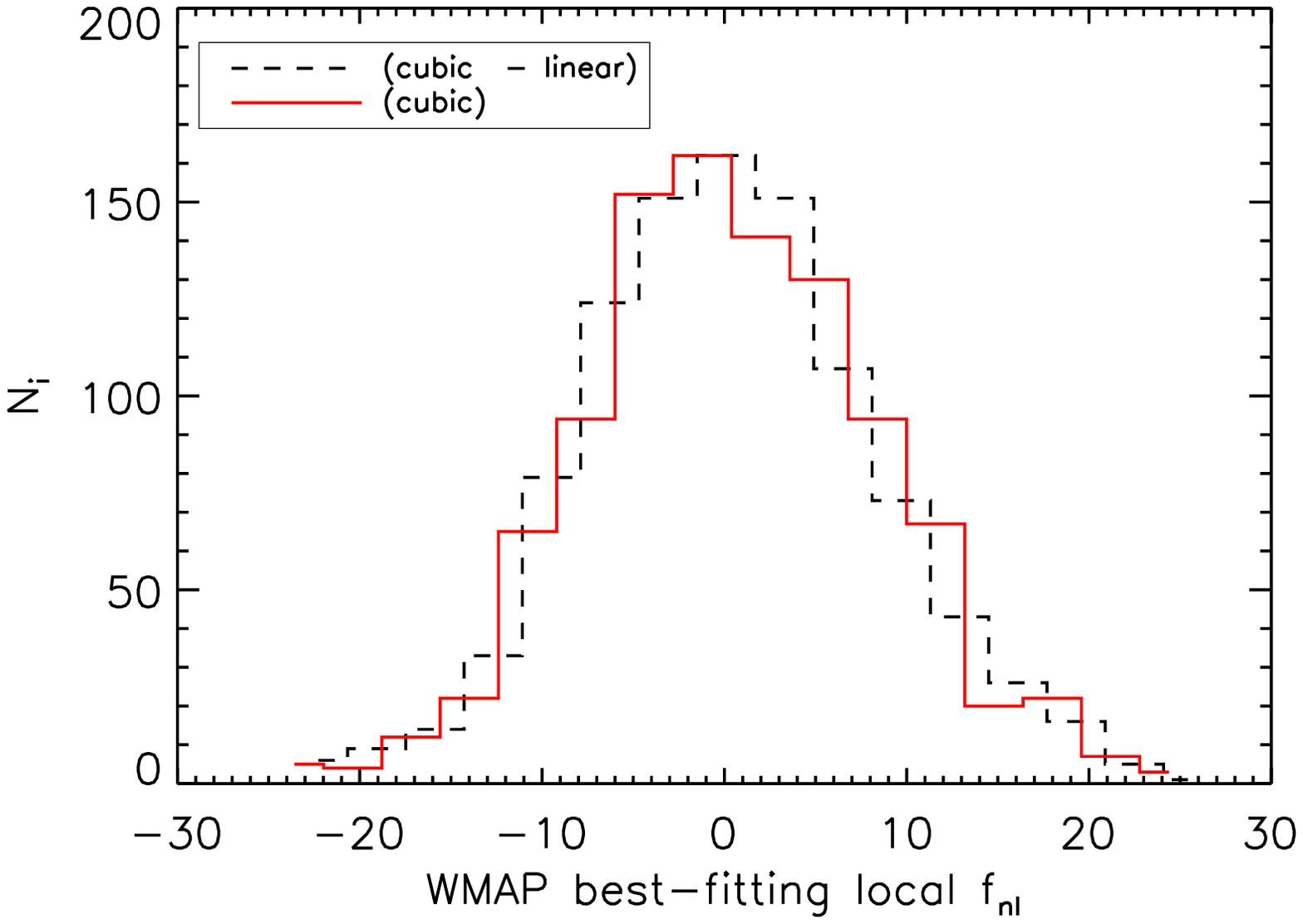}
\includegraphics[width=7.0cm, height=4.4cm, angle= 0]{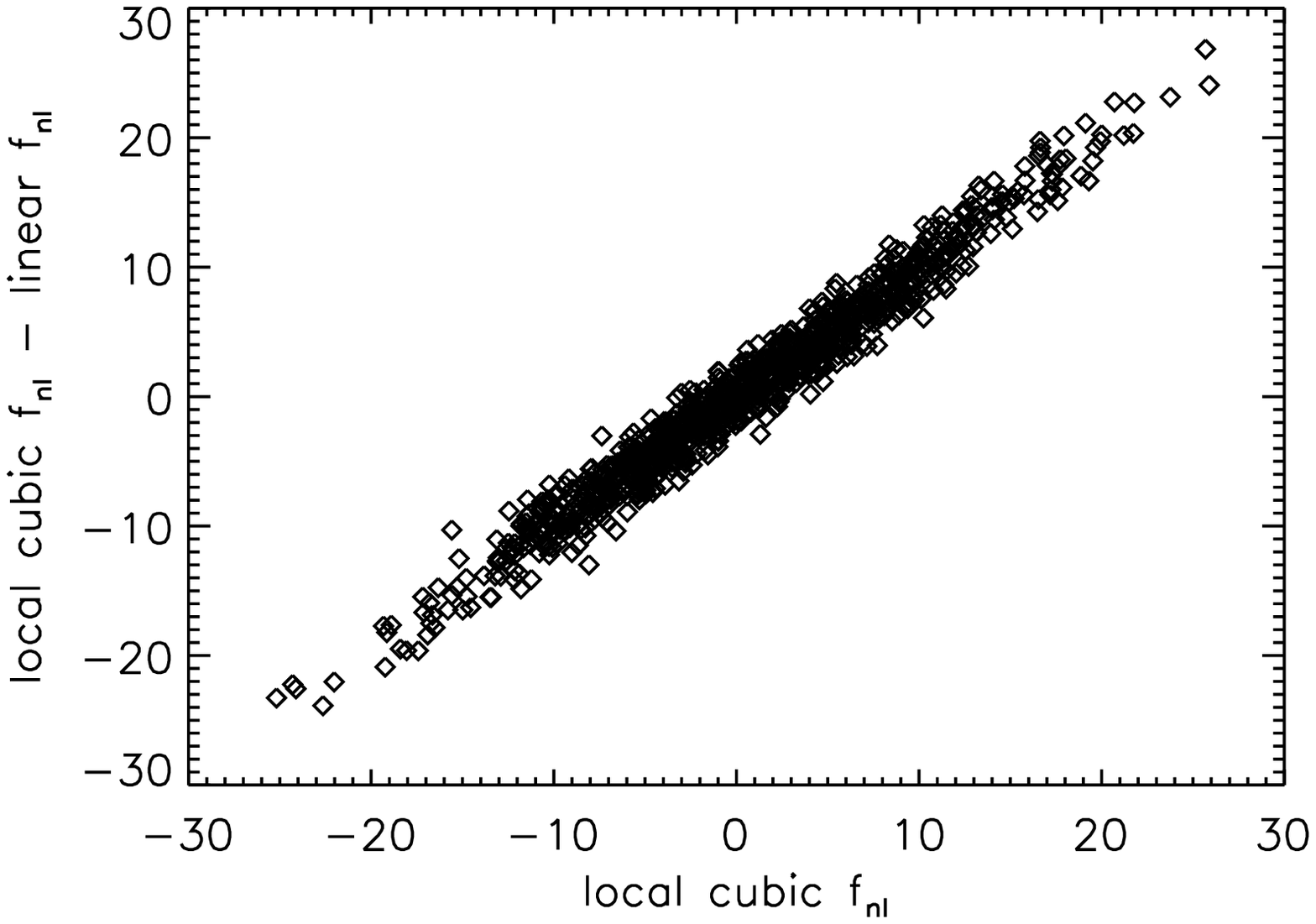}
\caption{ Expected Planck 143 GHz best-fitting $f_{nl}$ values
  using the cubic estimator (Eq. \ref{estimator_nolinearterm}) and the
  estimator with the linear correction
  (Eq. \ref{estimator_linear_term}) for the local shape. In the right
  panel the best-fitting $f_{nl}$ values with and without the linear
  term correction for each simulation are plotted. }
\label{fig:cubic_linear_fnl_planck}       
\end{figure*}

The cubic covariance matrix and the linear correlation matrices needed
for the $ f_{nl}$ estimator in Eq. (\ref{estimator_linear_term}) have
been computed using two independent sets of 10000 Planck Gaussian
simulations. The results corresponding to the analysis of an
additional set of 1000 Gaussian maps are presented in
Fig. \ref{fig:cubic_linear_fnl_planck}. Note that for this simulated
Planck level of anisotropy, $f^{(cubic)}_{nl}$ and $f^{(total)}_{nl}$
are also highly correlated. Finally in Table
\ref{table_fnl_results_local_planck} the properties of the previous
histograms are summarised. In particular, we see that using the cubic
estimator $\sigma(f_{nl}) = 7.98$ and the linear term contribution
reduces this error bar to $\sigma(f_{nl}) = 7.95$ (i.e. a negligible
correction lower than 0.4 per cent).
\begin{table}
  \caption{ Expected constraints on the $f_{nl}$ parameter for the
    local shape with and without the linear term correction for the
    143 GHz Planck channel. From left to right, the mean, dispersion,
    16, 84, 2.5 and 97.5 per cent quantiles respectively of the
    $f_{nl}$ distribution obtained with 1000 Gaussian maps. The Fisher
    error bar obtained for this case is $\sigma_F(f_{nl})=7.5$.}
\label{table_fnl_results_local_planck}
\begin{tabular}{c|cccccc}
\hline
case &  $\langle f_{nl} \rangle$ & $\sigma(f_{nl})$ & $X_{16}$ & $X_{84}$ & $X_{2.5}$ & $X_{97.5}$ \\
\hline
cubic & 0.4 & {\bf 7.98} & -7.2 & 8.6 & -15.0& 17.2 \\
linear & 0.0 & {\bf 1.43} & -1.4 & 1.5 & -2.7 & 2.8 \\
cubic - linear & 0.4 & {\bf 7.95} & -7.4 & 8.2 & -15.4 & 16.7 \\
\hline
\end{tabular}
\end{table}
%
\section{Conclusions}
\label{conclusions}
In this paper we have performed a comprehensive study of the $f_{nl}$
wavelet estimator. We have considered the main statistical assumptions
to derive the estimator in terms of the cubic quantities $q_i$ and
showed the conditions to reach optimality. We have found that the SMHW
wavelet produces an important effective decorrelation of the signal at
angular distances above the angular resolution $R$. This means that
the cubic quantities are nearly Gaussian owing to the central limit
theorem and using this property we have found an expression for the
likelihood of the $f_{nl}$ parameter in terms of the cubic statistics.
We have also included a linear term correction following the Wick
polynomials introduced by \citet{donzelli2012}. In particular, we have
confirmed that the linear term correction is basically achieved
through the mean subtraction that we carry out on the wavelet
coefficient maps for each angular scale
\citep{curto2011a,curto2011b}. We find that, in this case, the linear
term correction only reduces the error bars about 1 per cent for the
local case using WMAP data. This correction is even smaller for the
equilateral and orthogonal cases (0.2 per cent and 0.1 per cent
respectively). The results presented in this paper are in agreement
with the optimal results obtained with the wavelet estimator already
published where the mean subtraction was performed
\citep{curto2009a,curto2009b,curto2010,curto2011a,curto2011b}. Therefore,
we conclude that the contribution of the linear term is negligible
($\le$ 1 per cent) for the SMHW estimator for the three considered
shapes.  We have also explored the linear term correction for Planck
simulations at the 143 GHz channel. Our results indicate that the
correction for the local shape is lower than 0.4 per cent considering
the expected levels of noise anisotropy for this channel and the WMAP
KQ75 mask. From the results on WMAP data, we expect the correction for
the equilateral and orthogonal shapes to be even smaller.
\section*{acknowledgments}
The authors thank Biuse Casaponsa, Simona Donzelli, Michele Liguori,
Domenico Marinucci, Sabino Matarrese and Patricio Vielva for useful
comments. The authors acknowledge partial financial support from the
Spanish Ministerio de Econom\'ia y Competitividad project
AYA2010-21766-C03-01 and the Consolider Ingenio-2010 Programme project
CSD2010-00064. The authors acknowledge the computer resources,
technical expertise and assistance provided by the Spanish
Supercomputing Network (RES) node at Universidad de Cantabria.  We
also acknowledge the use of LAMBDA, support for which is provided by
the NASA Office of Space Science. The work has also used the software
package HEALPix \citep{healpix}. We acknowledge the use of the
pre-launch Planck Sky Model simulation package
\citep{delabrouille2012}.
%
%

\begin{thebibliography}{}

\bibitem[\protect\citeauthoryear{{Antoine} \& {Vandergheynst}}{{Antoine} \&
  {Vandergheynst}}{1998}]{antoine1998}
{Antoine} J.-P.,  {Vandergheynst} P.,  1998, Journal of Mathematical Physics,
  39, 3987

\bibitem[\protect\citeauthoryear{{Babich}}{{Babich}}{2005}]{babich2005}
{Babich} D.,  2005, Phys. Rev. D., 72, 043003

\bibitem[\protect\citeauthoryear{{Babich}, {Creminelli} \&
  {Zaldarriaga}}{{Babich} et~al.}{2004}]{babich2004}
{Babich} D.,  {Creminelli} P.,    {Zaldarriaga} M., , 2004, {The shape of
  non-Gaussianities}

\bibitem[\protect\citeauthoryear{{Bartolo}, {Komatsu}, {Matarrese} \&
  {Riotto}}{{Bartolo} et~al.}{2004}]{bartolo2004}
{Bartolo} N.,  {Komatsu} E.,  {Matarrese} S.,    {Riotto} A.,  2004, Phys.
  Rep., 402, 103

\bibitem[\protect\citeauthoryear{{Bucher}, {van Tent} \& {Carvalho}}{{Bucher}
  et~al.}{2010}]{bucher2010}
{Bucher} M.,  {van Tent} B.,    {Carvalho} C.~S.,  2010, MNRAS, 407, 2193

\bibitem[\protect\citeauthoryear{{Casaponsa}, {Barreiro}, {Curto},
  {Mart{\'{\i}}nez-Gonz{\'a}lez} \& {Vielva}}{{Casaponsa}
  et~al.}{2011a}]{casaponsa2011a}
{Casaponsa} B.,  {Barreiro} R.~B.,  {Curto} A.,  {Mart{\'{\i}}nez-Gonz{\'a}lez}
  E.,    {Vielva} P.,  2011a, MNRAS, 411, 2019

\bibitem[\protect\citeauthoryear{{Casaponsa}, {Bridges}, {Curto}, {Barreiro} \&
  {{Hobson}, M.~P. and Mart{\'{\i}}nez-Gonz{\'a}lez}}{{Casaponsa}
  et~al.}{2011b}]{casaponsa2011b}
{Casaponsa} B.,  {Bridges} M.,  {Curto} A.,  {Barreiro} R.~B.,    {{Hobson},
  M.~P. and Mart{\'{\i}}nez-Gonz{\'a}lez} E.,  2011b, MNRAS

\bibitem[\protect\citeauthoryear{{Cay{\'o}n}, {Mart{\'{\i}}nez-Gonz{\'a}lez},
  {Arg{\"u}eso}, {Banday} \& {G{\'o}rski}}{{Cay{\'o}n}
  et~al.}{2003}]{cayon2003}
{Cay{\'o}n} L.,  {Mart{\'{\i}}nez-Gonz{\'a}lez} E.,  {Arg{\"u}eso} F.,
  {Banday} A.~J.,    {G{\'o}rski} K.~M.,  2003, MNRAS, 339, 1189

\bibitem[\protect\citeauthoryear{{Creminelli}, {Nicolis}, {Senatore}, {Tegmark}
  \& {Zaldarriaga}}{{Creminelli} et~al.}{2006}]{creminelli2006}
{Creminelli} P.,  {Nicolis} A.,  {Senatore} L.,  {Tegmark} M.,    {Zaldarriaga}
  M.,  2006, Journal of Cosmology and Astro-Particle Physics, 5, 4

\bibitem[\protect\citeauthoryear{{Creminelli}, {Senatore} \&
  {Zaldarriaga}}{{Creminelli} et~al.}{2007}]{creminelli2007}
{Creminelli} P.,  {Senatore} L.,    {Zaldarriaga} M.,  2007, Journal of
  Cosmology and Astro-Particle Physics, 3, 19

\bibitem[\protect\citeauthoryear{{Curto}, {Mart{\'{\i}}nez-Gonz{\'a}lez} \&
  {Barreiro}}{{Curto} et~al.}{2009b}]{curto2009b}
{Curto} A.,  {Mart{\'{\i}}nez-Gonz{\'a}lez} E.,    {Barreiro} R.~B.,  2009b,
  ApJ, 706, 399

\bibitem[\protect\citeauthoryear{{Curto}, {Mart{\'{\i}}nez-Gonz{\'a}lez} \&
  {Barreiro}}{{Curto} et~al.}{2010}]{curto2010}
{Curto} A.,  {Mart{\'{\i}}nez-Gonz{\'a}lez} E.,    {Barreiro} R.~B.,  2010, in
  {J.~M.~Diego, L.~J.~Goicoechea, J.~I.~Gonz{\'a}lez-Serrano, \& J.~Gorgas}
  ed., Highlights of Spanish Astrophysics V {Constraints on the Non-linear
  Coupling Parameter $f _{nl}$ Using the CMB}.
pp 277--+

\bibitem[\protect\citeauthoryear{{Curto}, {Mart{\'{\i}}nez-Gonz{\'a}lez} \&
  {Barreiro}}{{Curto} et~al.}{2011a}]{curto2011a}
{Curto} A.,  {Mart{\'{\i}}nez-Gonz{\'a}lez} E.,    {Barreiro} R.~B.,  2011a,
  MNRAS, 412, 1038

\bibitem[\protect\citeauthoryear{{Curto}, {Mart{\'{\i}}nez-Gonz{\'a}lez},
  {Barreiro} \& {Hobson}}{{Curto} et~al.}{2011b}]{curto2011b}
{Curto} A.,  {Mart{\'{\i}}nez-Gonz{\'a}lez} E.,  {Barreiro} R.~B.,    {Hobson}
  M.~P.,  2011b, MNRAS, 417, 488

\bibitem[\protect\citeauthoryear{{Curto}, {Mart{\'{\i}}nez-Gonz{\'a}lez},
  {Mukherjee}, {Barreiro}, {Hansen}, {Liguori} \& {Matarrese}}{{Curto}
  et~al.}{2009a}]{curto2009a}
{Curto} A.,  {Mart{\'{\i}}nez-Gonz{\'a}lez} E.,  {Mukherjee} P.,  {Barreiro}
  R.~B.,  {Hansen} F.~K.,  {Liguori} M.,    {Matarrese} S.,  2009a, MNRAS, 393,
  615

\bibitem[\protect\citeauthoryear{{Delabrouille}, {Betoule}, {Melin},
  {Miville-Desch\^enes}, {Gonz\'alez-Nuevo} \& {et al.}}{{Delabrouille}
  et~al.}{2012}]{delabrouille2012}
{Delabrouille} J.,  {Betoule} M.,  {Melin} J.~B.,  {Miville-Desch\^enes} M.~A.,
   {Gonz\'alez-Nuevo} J., {et al.} 2012, in preparation

\bibitem[\protect\citeauthoryear{{Donzelli}, {Hansen}, {Liguori}, {Marinucci}
  \& {Matarrese}}{{Donzelli} et~al.}{2012}]{donzelli2012}
{Donzelli} S.,  {Hansen} F.~K.,  {Liguori} M.,  {Marinucci} D.,    {Matarrese}
  S.,  2012, preprint (arXiv:1202.1478)

\bibitem[\protect\citeauthoryear{{D}, {Hansen}, {Liguori}, {Marinucci}
  \& {Matarrese}}{{Donzelli} et~al.}{2012}]{donzelli2012}
{Donzelli} S.,  {Hansen} F.~K.,  {Liguori} M.,  {Marinucci} D.,    {Matarrese}
  S.,  2012, preprint (arXiv:1202.1478)

\bibitem[\protect\citeauthoryear{{Elsner} \& {Wandelt}}{{Elsner} \&
  {Wandelt}}{2009}]{elsner2009}
{Elsner} F.,  {Wandelt} B.~D.,  2009, ApJS, 184, 264

\bibitem[\protect\citeauthoryear{{Elsner} \& {Wandelt}}{{Elsner} \&
  {Wandelt}}{2010}]{elsner2010b}
{Elsner} F.,  {Wandelt} B.~D.,  2010, ApJ, 724, 1262

\bibitem[\protect\citeauthoryear{{Elsner}, {Wandelt} \& {Schneider}}{{Elsner}
  et~al.}{2010}]{elsner2010a}
{Elsner} F.,  {Wandelt} B.~D.,    {Schneider} M.~D.,  2010, A\&A, 513, A59+

\bibitem[\protect\citeauthoryear{{Fergusson} \& {Shellard}}{{Fergusson} \&
  {Shellard}}{2011}]{fergusson2011}
{Fergusson} J.,  {Shellard} E.~S.,  2011, preprint (arXiv:1105.2791)

\bibitem[\protect\citeauthoryear{{Fergusson}, {Liguori} \&
  {Shellard}}{{Fergusson} et~al.}{2010a}]{fergusson2010a}
{Fergusson} J.~R.,  {Liguori} M.,    {Shellard} E.~P.~S.,  2010a, Phys. Rev. D,
  82, 023502

\bibitem[\protect\citeauthoryear{{Fergusson}, {Liguori} \&
  {Shellard}}{{Fergusson} et~al.}{2010b}]{fergusson2010b}
{Fergusson} J.~R.,  {Liguori} M.,    {Shellard} E.~P.~S.,  2010b, preprint (arXiv:1006.1642)

\bibitem[\protect\citeauthoryear{{G{\'o}rski}, {Hivon}, {Banday}, {Wandelt},
  {Hansen}, {Reinecke} \& {Bartelmann}}{{G{\'o}rski} et~al.}{2005}]{healpix}
{G{\'o}rski} K.~M.,  {Hivon} E.,  {Banday} A.~J.,  {Wandelt} B.~D.,  {Hansen}
  F.~K.,  {Reinecke} M.,    {Bartelmann} M.,  2005, ApJ, 622, 759

\bibitem[\protect\citeauthoryear{{Komatsu}, {Dunkley}, {Nolta}, {Bennett},
  {Gold}, {Hinshaw}, {Jarosik}, {Larson}, {Limon}, {Page}, {Spergel},
  {Halpern}, {Hill}, {Kogut}, {Meyer}, {Tucker}, {Weiland}, {Wollack} \&
  {Wright}}{{Komatsu} et~al.}{2009}]{komatsu2009}
{Komatsu} E.,  {Dunkley} J.,  {Nolta} M.~R.,  {Bennett} C.~L.,  {Gold} B.,
  {Hinshaw} G.,  {Jarosik} N.,  {Larson} D.,  {Limon} M.,  {Page} L.,
  {Spergel} D.~N.,  {Halpern} M.,  {Hill} R.~S.,  {Kogut} A.,  {Meyer} S.~S.,
  {Tucker} G.~S.,  {Weiland} J.~L.,  {Wollack} E.,    {Wright} E.~L.,  2009,
  ApJS, 180, 330

\bibitem[\protect\citeauthoryear{{Komatsu}, {Kogut}, {Nolta}, {Bennett},
  {Halpern}, {Hinshaw}, {Jarosik}, {Limon}, {Meyer}, {Page}, {Spergel},
  {Tucker}, {Verde}, {Wollack} \& {Wright}}{{Komatsu}
  et~al.}{2003}]{komatsu2003}
{Komatsu} E.,  {Kogut} A.,  {Nolta} M.~R.,  {Bennett} C.~L.,  {Halpern} M.,
  {Hinshaw} G.,  {Jarosik} N.,  {Limon} M.,  {Meyer} S.~S.,  {Page} L.,
  {Spergel} D.~N.,  {Tucker} G.~S.,  {Verde} L.,  {Wollack} E.,    {Wright}
  E.~L.,  2003, ApJS, 148, 119

\bibitem[\protect\citeauthoryear{{Komatsu}, {Smith}, {Dunkley}, {Bennett},
  {Gold} \& {et al.}}{{Komatsu} et~al.}{2011}]{komatsu2011}
{Komatsu} E.,  {Smith} K.~M.,  {Dunkley} J.,  {Bennett} C.~L.,  {Gold} B.,
  {et al.} 2011, ApJS, 192, 18

\bibitem[\protect\citeauthoryear{{Komatsu} \& {Spergel}}{{Komatsu} \&
  {Spergel}}{2001}]{komatsu2001}
{Komatsu} E.,  {Spergel} D.~N.,  2001, Phys. Rev. D, 63, 063002

\bibitem[\protect\citeauthoryear{{Komatsu}, {Spergel} \& {Wandelt}}{{Komatsu}
  et~al.}{2005}]{komatsu2005}
{Komatsu} E.,  {Spergel} D.~N.,    {Wandelt} B.~D.,  2005, ApJ, 634, 14

\bibitem[\protect\citeauthoryear{{Komatsu}, {Wandelt}, {Spergel}, {Banday} \&
  {G{\'o}rski}}{{Komatsu} et~al.}{2002}]{komatsu2002}
{Komatsu} E.,  {Wandelt} B.~D.,  {Spergel} D.~N.,  {Banday} A.~J.,
  {G{\'o}rski} K.~M.,  2002, ApJ, 566, 19

\bibitem[\protect\citeauthoryear{{Liguori}, {Sefusatti}, {Fergusson} \&
  {Shellard}}{{Liguori} et~al.}{2010}]{liguori2010}
{Liguori} M.,  {Sefusatti} E.,  {Fergusson} J.~R.,    {Shellard} E.~P.~S.,
  2010, Advances in Astronomy, 2010

\bibitem[\protect\citeauthoryear{{Marinucci}, {Pietrobon}, {Balbi}, {Baldi},
  {Cabella}, {Kerkyacharian}, {Natoli}, {Picard} \& {Vittorio}}{{Marinucci}
  et~al.}{2008}]{marinucci2008}
{Marinucci} D.,  {Pietrobon} D.,  {Balbi} A.,  {Baldi} P.,  {Cabella} P.,
  {Kerkyacharian} G.,  {Natoli} P.,  {Picard} D.,    {Vittorio} N.,  2008,
  MNRAS, 383, 539

\bibitem[\protect\citeauthoryear{{Mart{\'{\i}}nez-Gonz{\'a}lez}}{{Mart{\'{\i}}%
nez-Gonz{\'a}lez}}{2008}]{martinez2008}
{Mart{\'{\i}}nez-Gonz{\'a}lez} E.,  2008, preprint (arXiv:0805.4157)

\bibitem[\protect\citeauthoryear{{Mart{\'{\i}}nez-Gonz{\'a}lez}, {Gallegos},
  {Arg{\"u}eso}, {Cay{\'o}n} \& {Sanz}}{{Mart{\'{\i}}nez-Gonz{\'a}lez}
  et~al.}{2002}]{martinez2002}
{Mart{\'{\i}}nez-Gonz{\'a}lez} E.,  {Gallegos} J.~E.,  {Arg{\"u}eso} F.,
  {Cay{\'o}n} L.,    {Sanz} J.~L.,  2002, MNRAS, 336, 22

\bibitem[\protect\citeauthoryear{{McEwen}, {Vielva}, {Wiaux}, {Barreiro},
  {Cayon}, {Hobson}, {Lasenby}, {Martinez-Gonzalez} \& {Sanz}}{{McEwen}
  et~al.}{2007}]{mcewen2007}
{McEwen} J.~D.,  {Vielva} P.,  {Wiaux} Y.,  {Barreiro} R.~B.,  {Cayon} L.,
  {Hobson} M.~P.,  {Lasenby} A.~N.,  {Martinez-Gonzalez} E.,    {Sanz} J.~L.,
  2007, Journal of Fourier Analysis and Applications, 13, 495

\bibitem[\protect\citeauthoryear{{Pietrobon}, {Cabella}, {Balbi}, {de Gasperis}
  \& {Vittorio}}{{Pietrobon} et~al.}{2009}]{pietrobon2009}
{Pietrobon} D.,  {Cabella} P.,  {Balbi} A.,  {de Gasperis} G.,    {Vittorio}
  N.,  2009, MNRAS, 396, 1682

\bibitem[\protect\citeauthoryear{{Rudjord}, {Hansen}, {Lan}, {Liguori},
  {Marinucci} \& {Matarrese}}{{Rudjord} et~al.}{2009}]{rudjord2009}
{Rudjord} {\O}.,  {Hansen} F.~K.,  {Lan} X.,  {Liguori} M.,  {Marinucci} D.,
  {Matarrese} S.,  2009, ApJ, 701, 369

\bibitem[\protect\citeauthoryear{{Sanz}, {Herranz}, {L\'opez-Caniego} \&
  {Arg\"ueso}}{{Sanz} et~al.}{2006}]{sanz2006}
{Sanz} J.~L.,  {Herranz} D.,  {L\'opez-Caniego} M.,    {Arg\"ueso} F.,  2006, preprint (arXiv:astro-ph/0609351)

\bibitem[\protect\citeauthoryear{{Senatore}, {Smith} \&
  {Zaldarriaga}}{{Senatore} et~al.}{2010}]{senatore2010}
{Senatore} L.,  {Smith} K.~M.,    {Zaldarriaga} M.,  2010, Journal of Cosmology
  and Astro-Particle Physics, 1, 28

\bibitem[\protect\citeauthoryear{{Smidt}, {Amblard}, {Byrnes}, {Cooray},
  {Heavens} \& {Munshi}}{{Smidt} et~al.}{2010}]{smidt2010}
{Smidt} J.,  {Amblard} A.,  {Byrnes} C.~T.,  {Cooray} A.,  {Heavens} A.,
  {Munshi} D.,  2010, Phys. Rev. D, 81, 123007

\bibitem[\protect\citeauthoryear{{Smith}, {Senatore} \& {Zaldarriaga}}{{Smith}
  et~al.}{2009}]{smith2009}
{Smith} K.~M.,  {Senatore} L.,    {Zaldarriaga} M.,  2009, Journal of Cosmology
  and Astro-Particle Physics, 9, 6

\bibitem[\protect\citeauthoryear{{Spergel}, {Bean}, {Dor{\'e}}, {Nolta},
  {Bennett}, {Dunkley}, {Hinshaw}, {Jarosik} \& {et al.}}{{Spergel}
  et~al.}{2007}]{spergel2007}
{Spergel} D.~N.,  {Bean} R.,  {Dor{\'e}} O.,  {Nolta} M.~R.,  {Bennett} C.~L.,
  {Dunkley} J.,  {Hinshaw} G.,  {Jarosik} N.,    {et al.} 2007, ApJS, 170, 377

\bibitem[\protect\citeauthoryear{{Vielva}, {Mart{\'{\i}}nez-Gonz{\'a}lez} \&
  {Tucci}}{{Vielva} et~al.}{2006}]{vielva2006}
{Vielva} P.,  {Mart{\'{\i}}nez-Gonz{\'a}lez} E.,    {Tucci} M.,  2006, MNRAS,
  365, 891

\bibitem[\protect\citeauthoryear{{Vielva}}{{Vielva}}{2007}]{vielva2007b}
{Vielva} P.,  2007, in Society of Photo-Optical Instrumentation Engineers
  (SPIE) Conference Series Vol.~6701 of Society of Photo-Optical
  Instrumentation Engineers (SPIE) Conference Series, {Probing the Gaussianity
  and the statistical isotropy of the CMB with spherical wavelets}

\bibitem[\protect\citeauthoryear{{Yadav} \& {Wandelt}}{{Yadav} \&
  {Wandelt}}{2008}]{wandelt2008}
{Yadav} A.~P.~S.,  {Wandelt} B.~D.,  2008, Physical Review Letters, 100, 181301

\bibitem[\protect\citeauthoryear{{Yadav} \& {Wandelt}}{{Yadav} \&
  {Wandelt}}{2010}]{yadav2010}
{Yadav} A.~P.~S.,  {Wandelt} B.~D.,  2010, Advances in Astronomy, 2010

\bibitem[\protect\citeauthoryear{{Yu}, {Harnois-D{\'e}raps}, {Zhang} \&
  {Pen}}{{Yu} et~al.}{2012}]{yu2012}
{Yu} H.-R.,  {Harnois-D{\'e}raps} J.,  {Zhang} T.-J.,    {Pen} U.-L.,  2012,
  MNRAS, 421, 832

\bibitem[\protect\citeauthoryear{{Zhang}, {Yu}, {Harnois-D{\'e}raps},
  {MacDonald} \& {Pen}}{{Zhang} et~al.}{2011}]{zang2011}
{Zhang} T.-J.,  {Yu} H.-R.,  {Harnois-D{\'e}raps} J.,  {MacDonald} I.,    {Pen}
  U.-L.,  2011, ApJ, 728, 35

\end{thebibliography}

%
\end{document}